\documentclass[11pt]{article} 
\usepackage[utf8]{inputenc}

\usepackage{amsfonts}
\usepackage{amsmath}
\usepackage{amssymb}
\usepackage{amsthm}
\usepackage{caption}
\usepackage{color}
    \definecolor{darkgreen}{rgb}{0,0.5,0}
    \definecolor{darkblue}{rgb}{0,0,0.6}
    \definecolor{purple}{rgb}{0.4,.2,0.7}
\usepackage[margin = 2.5cm]{geometry}
\usepackage{graphicx}
\usepackage[hyperfootnotes = false, colorlinks = true, linkcolor = darkblue, citecolor = purple]{hyperref}
\usepackage{subcaption}

\def \beq { \begin{equation}}
\def \eeq {\end{equation}}

\def \al {\alpha}

\def \om {\omega}

\def \ep {\epsilon}

\def \Lc {\mathcal{L}}

\def \pr {\partial}
\def \ra {\rightarrow}

\def \oh {\cfrac{1}{2}}

\def \beq { \begin{equation}}
\def \eeq {\end{equation}}

\DeclareMathOperator*{\Tr}{Tr}

\renewcommand\Re{\operatorname{Re}}
\renewcommand\Im{\operatorname{Im}}
\newcommand\const{\operatorname{const}}

\newcommand\sgn{\operatorname{sgn}}

\def \l {\left(}
\def \r {\right)}

\def \bra {\langle}
\def \ket {\rangle}
\def \oh {\frac{1}{2}}
\def \Oc {\mathcal{O}}
\def \Lc {\mathcal{L}}
\def \Jc {\mathcal{J}}
\def \Cc {\mathcal{C}}

\def \tt {\tilde{t}}
 
\def \tBH {\Delta t_{\rm 2BH}}
\def \tWH {\Delta t_{\rm WH}}

\newcommand{\be}{\begin{equation}}
\newcommand{\ee}{\end{equation}}
 \def\nref#1{(\ref{#1})}
\def \la {\label}   
\def \be {\begin{equation}}
\def \ee {\end{equation}}
\newcommand{\bea}{\begin{eqnarray}}
\newcommand{\eea}{\end{eqnarray}}
\def \half {{1\over 2}}	
\def \JM#1 {{\color{blue}  JM: #1 }}
 \def \AM#1 {{\color{red}  AM: #1 }}
 
\def\T2BH{T_{\rm 2BH}}
\def\TWH{T_{\rm WH}}
\def\bet2BH{\beta_{\rm 2 BH}}
\def\betWH{\beta_{\rm WH}}
\def\tpr{{\bf t}'}

\begin{document}

 \vskip 2cm
 
\title{\bf SYK wormhole formation in real time}
\vskip 4cm
\author{Juan~Maldacena$^1$ and Alexey~Milekhin$^2$}
\vskip 2cm
\date{$^1$ \it Institute for Advanced Study,  Princeton, NJ 08540, USA \\
$^2$\it Physics Department, Princeton University, Princeton, NJ 08544, USA\\
}
\maketitle
\vskip 3cm
\abstract{
We study the real time formation of the ground state of two coupled SYK models. 
This is a highly entangled state which is close to the thermofield double state and can be viewed as a wormhole. 
We start from a high temperature state, we let it cool by coupling to a cold bath. We numerically solve for the large $N$ dynamics.
Our main result is that the system  forms a wormhole by going through a region with negative specific heat, 
taking time that is independent of $N$.  The dynamics is smooth everywhere and  it seems to follow the equilibrium thermodynamic configurations of the microcanonical ensemble. 
Also we comment on the relation between this coupled  SYK model  and Jackiw--Teitelboim gravity theory with bulk fields. 
 }

\eject

\tableofcontents

\eject
\section{Introduction and Summary}

\subsection{Motivation} 

The Sachdev--Ye--Kitaev (SYK) model \cite{SachdevYe,kitaev,MS,KitaevSuhFirstPaper} is a strongly interacting but yet solvable model in the large $N$ limit. At low energies, it displays an approximate conformal symmetry. In this region, the model has many features in common with nearly $AdS_2$ gravity, or Jackiw-Teitelboim (JT) gravity \cite{Jackiw,Teitelboim, AP}, coupled to matter fields. This is a simple two dimensional theory of gravity which describes some aspects of nearly extremal black holes in various dimensions. 

An interesting variant is to consider a pair of identical SYK models coupled through a simple bilinear interaction \cite{MQ}, see also 
\cite{Garcia-Garcia:2019poj,Kim:2019upg,GaoJafferisRecent}. 
The ground state of this model has a gap, but its excitation spectrum also displays an approximate conformal symmetry. Furthermore, this ground state is close to the thermofield double state of two decoupled models. For reasons we explain below,  we call the ground state of this coupled model  ``the SYK wormhole''. 

A conceptually similar state also arises when one considers two nearly extremal black holes that are relatively close, so that they are coupled. In this case, a traversable wormhole can connect the near extremal throats \cite{MMP}. 
This can be effectively modeled by a nearly $AdS_2$ gravity theory 
where we have  direct  interactions between the values of the bulk fields near the two boundaries \cite{GaoJafferisWall,MQ}. In other words, thinking of the Penrose diagram of $AdS_2$ as a strip, we put boundary conditions for the bulk fields that connect the two boundaries. 
The two boundaries are causally connected through the bulk, so that this spacetime describes a wormhole. 
 This wormhole is the lowest energy configuration of the system and it also displays the approximate $SL(2,R)$ isometries of nearly $AdS_2$. 
   
   Given that this is a remarkable state, we are interested in knowing whether it is easy to get to it. In other words, if we start out from a general excited state of the coupled model, can we easily get to the ground state by cooling the system down? Or will the system get stuck in some other state?  At first sight the answer seems straightforward, if it is the ground state, the system will surely find it if it can shed its excess energy to the bath. On the other hand, from the gravity perspective, the process involves a topology change. Such topology change might happen via a tunneling solution, but it would be exponentially suppressed in $N$ (or the entropy of each separate black hole). 
  
  \subsection{Wormhole formation in SYK} 
   
  With this motivation in mind, we study this problem for the 
  two coupled SYK models. We start with a relatively high temperature state of the coupled model which looks like two thermal density matrices, one for each SYK factor. Then we couple the system to a bath and study the evolution in real time by solving the large $N$ Schwinger-Dyson equations. We find that the system indeed finds the ``SYK wormhole" ground state in a time that is independent of $N$. In particular, there is no exponential suppression.  
  Notice that the ability to efficiently find this ground state also makes it possible to prepare the thermofield double (TFD) state of the decoupled model, by simply switching off the interaction between the two sides \cite{MQ}, after we have found the ground state.   

The approach we used is the following. The large $N$ 
Dyson--Schwinger equations  form a closed system for the
two-point function \cite{SachdevYe,kitaev,MS,KitaevSuhFirstPaper}.
 In the out-of-equilibrium situation that we are considering,
 these equations are commonly referred to as the Kadanoff--Baym (KB) equations.   We couple the system of two interacting SYK models to a large bath and find the real 
 time dynamics using KB equations.
The problem of coupling SYK to a bath was recently studied in \cite{syk_bath} and we borrow some results from there. Also, the KB equations for a single SYK
was recently studied numerically in \cite{quench1,quench2}. 
Here we study this problem solving the dynamical equations at $q=4$.
 The problem has many time scales and due to numerical limitations we could not separate them all by large amounts. However, 
 our numerical results seem to confirm the picture where the system follows the microcanonical equilibrium curve.
  We now briefly review this equilibrium thermodynamics. 


\subsection{Equilibrium thermodynamics} 

In the canonical ensemble the system has two phases: the low temperature one corresponding to 
the ground state, the SYK wormhole,  and its excitations; and a higher temperature phase which is closer to two separate thermal SYK systems. 
The two phases are separated by a first order phase transition. 
In the large $q$ limit,  
the black hole phase and the wormhole phase are   smoothly connected by a canonically unstable  phase with negative specific heat \cite{MQ}. However,   
in the microcanonical ensemble, we expect that the system  smoothly interpolates between these two phases. In other words, in the microcanonical ensemble we expect no phase transition as we lower the energy.

\begin{figure}[!ht]
\begin{center}
\includegraphics[scale=0.5]{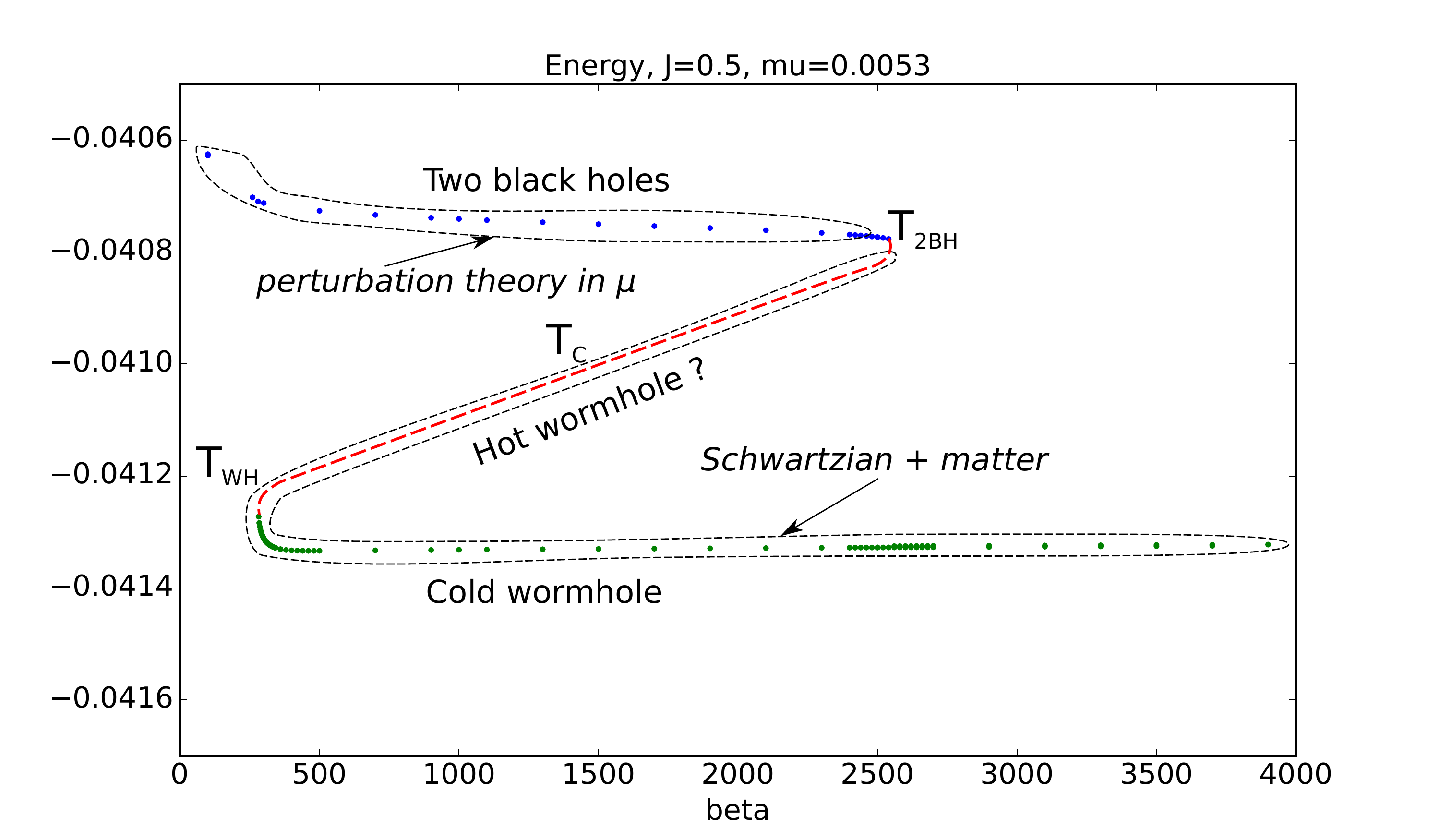}
\minipage{0.8\textwidth}
\caption{ \small
Dots: energy vs beta obtained by numerically solving DS equations for two coupled SYK models with $J=0.5, \mu=0.0053$.
Blue dots correspond to the ``two black holes phase'', whereas green dots correspond to the ``cold wormhole phase''.
Red dashed line:  curve for the ``hot wormhole'' phase expected from a low energy analytic analysis.   
The question mark ``?'' reminds us that we were not able to find it as a solution of the euclidean DS equations. }
\label{fig:energy_fancy}
\endminipage
\end{center}
\end{figure}
Figure \ref{fig:energy_fancy} shows energy vs inverse temperature $\beta$ for $q=4$. We use energy instead of free energy or entropy because   we will be dealing
mostly with Lorentzian non-equilibrium correlators numerically and it is easier to find the temperature and the energy from them.
There are three different regions. At high temperatures $T>\T2BH$ we have the phase we name the ``two black holes phase''.
  At low temperatures $T<\TWH$ we have  the phase we call ``cold wormhole" phase, which can be viewed a as a wormhole with few thermal excitations.  The two phases overlap, since $\T2BH < \TWH$.  In the intermediate
temperatures regime $\T2BH < T < \TWH$ we also expect a canonically unstable, but microcanonically stable, phase that we call 
``the hot wormhole phase''.
  As we mentioned above, 
this phase can be found analytically in the large $q$ limit. It   has positive entropy but negative specific heat. 
However, at finite $q$ we do not known much about this phase, since we have been unable to find it by solving the Euclidean Schwinger Dyson equations. We interpret this failure as resulting from its canonical instability. This is why we put a question mark in Figure \ref{fig:energy_fancy}.   We will find evidence for this phase through the real time evolution, since we will find that the temperature goes up as the energy monotonically goes down.  There is also analytic evidence from a low energy analysis, as we will review later. 
The names ``hot'' or ``cold'' wormhole refer to how these would feel to an observer who is inside the wormhole, at its center, in a gravity picture for these configurations.  
As is clear from Figure \ref{fig:energy_fancy}, there are outside temperatures where we can have both a ``cold'' and ``hot'' wormhole. When we talk about temperatures in this paper, we are always referring to the physical temperature as seen from the outside.  

Figure \ref{fig:energy_fancy} also displays the critical temperature, $T_c$,   when two stable phases switch dominance in the canonical ensemble. 
For $\T2BH < T <T_c$ two black hole phase is thermodynamically metastable but is not a global minimum for the free energy. Similarly for 
$T_c < T < \TWH$ and the cold wormhole phase. For our problem the microcanonical ensemble is more relevant. Notice that the different ``phases'' are continuously connected in the microcanonical ensemble, so they are not really sharply separated  phases. 


For small values of $\mu$, the coupling between the two SYK models, 
 we can make different analytic approximations for the different parts of the curve.  
For the two black hole region we can start with two separate thermal SYK models and use  
perturbation theory in  
$\mu$. 
 In this regime, the left-right correlator $G_{LR}$ is small and   of order $\mu$. The gravity picture is that we have two separate hyperbolic disks with a boundary perturbation that connects the bulk  fields on the two disks. 
We find that $\T2BH$ is in the region where this perturbation theory breaks down. 
For low temperatures the left-right correlator is of order one. We can access this regime by assuming that the system is close
to the thermofield double state. The relevant part of the dynamics is captured by the Schwarzian mode. This aspect of the dynamics is the same for the SYK model and the nearly-$AdS_2$ gravity theory  \cite{cft_breaking,Jensen:2016pah,Engelsoy:2016xyb}. 
This describes both the cold wormhole and hot wormhole phases. In particular, we can see the existence of the hot wormhole phase in this approximation \cite{MQ}. 
In particular,    the temperature $\TWH$ can be found within this approximation. We review this description in Section \ref{sec:sch_low}.

\subsection{Gravity picture } 

One of our motivations was to understand whether a similar wormhole formation process occurs in more general theories of nearly-$AdS_2$ gravity with matter.  
  

With this goal in mind we will present a gravity picture for the transition we have in the SYK model. 
We do not know the precise gravity dual of the SYK model. But we consider a nearly-$AdS_2$ gravity theory that has some of the same features. For questions that mainly involve the Schwarzian mode,  the SYK and
nearly-$AdS_2$ answers 
match precisely \cite{cft_breaking,Jensen:2016pah,Engelsoy:2016xyb}. 
However,   wormhole formation goes slightly beyond this approximation,  and we need to incorporate one important feature that is related to the origin of the ground state entropy, or ``zero temperature'' entropy, 
$S_0$, of the SYK model. If we start from a phase consisting of two thermal states, then the entropy will have a large contribution of size
$2 S_0$ (plus thermal corrections). As the wormhole forms, the system should be able to shed this large entropy into the bath. In gravity this involves topology change, which would naively be exponentially suppressed. On the other hand, as we discussed above,  this happens without any such suppression in SYK.

 We can reproduce this from a gravity picture as follows. First we view the two coupled systems as a nearly-$AdS_2$ gravity theory with $N$ bulk fermion fields with Neumann boundary conditions. The two black hole phase consists of two hyperbolic disks with an interaction between the boundary values of the bulk fields. 
As we lower the temperature, this interaction effectively becomes strong and the theory flows to a new IR fixed point. The new fixed point is simply a theory with different boundary conditions, namely Dirichlet boundary conditions.  
This change in boundary conditions is similar to the one discussed in \cite{KlebanovWitten}.  
The two disks decouple again but the boundary conditions are effectively changed from Neumann to Dirichlet. Now we use the observation in \cite{KitaevRecent},  that $e^{S_0}$ is equal to the ratio of the Neumann vs Dirichlet  disk partition functions for $N$ fermions. This means that the effective theory in the IR, with Dirichlet boundary condition has now $S_0^{eff} =0$. This means that topology change ``costs us nothing'', and we can easily transition into the wormhole phase. In fact, by a similar argument we can say that the end of the hot wormhole phase also corresponds to the region where the interactions between the two sides of the global $AdS_2$ strip produce a flow that change the boundary conditions of the fermions from Neumann to Dirichlet. 

In summary, we provide a qualitative gravity mechanism for the formation of the SYK wormhole. The purpose of this explanation was to contrast SYK with what we expect in a generic gravity theory. 
A generic gravity theory can have a number of fields much smaller than  $S_0$. In this case, the change in boundary conditions would not significantly change $S_0$ and it would still be difficult to change the topology. 
 For this reason we could not answer the question of whether there is an 
 ``easy way'' of forming the wormhole for more 
general gravity theories, such as the case of four 
dimensional magnetically charged wormholes in the Standard Model \cite{MMP}.


The paper is organized as follows. In Section \ref{MQModel} we review the two coupled SYK model \cite{MQ}. We   describe the perturbative approach at high temperatures, for the ``two black hole phase''. We also review the Schwarzian description of the low energy dynamics that describes the hot and cold wormhole phases. 
Section \ref{RealTime} contains our real time analysis of the formation of the wormhole. We set up the coupling to the bath, we write the 
Kadanoff-Baym equations (the real time Schwinger Dyson equations), and we present the result of a   numerical analysis for some particular values of the parameters. 
In section \ref{sec:gravity}, we discuss
differences and similarities between SYK and nearly-$AdS_2$ (or JT) gravity and also provide the qualitative picture of the transition.
Various computational details are discussed in the appendices.

\section{The two coupled SYK model and its thermodynamics }
\label{MQModel}

\subsection{Definition and properties of the ground state}
\label{sec:mq}


Here we review properties of the two coupled SYK models introduced in \cite{MQ}. 
The Hamiltonian of the model consists of two SYK terms coupled by an interaction
\be \la{CoupMod}
H = H_{\rm SYK }^L + H_{\rm SYK}^R + H_{\rm int} 
\ee
where 
\beq
 H_{\rm SYK }^L + H_{\rm SYK}^R  =  \sum_{ 1 \le i < j < k < l \le  N }J_{ijkl} \l \psi_i^L \psi_j^L \psi_k^L \psi_l^L + 
\psi_i^R \psi_j^R \psi_k^R \psi_l^R \r
\eeq
where the couplings are the same for both factors. They are Gaussian random variables with variance 
\beq
\bra J_{ijkl}^2 \ket = \frac{3! J^2 }{N^3},~~ \ \text{no sum}
\eeq
There is also a generalization where we consider a $q$ fermion interaction term, 
 instead of four. 
The interaction term has the form 
\beq
\label{eq:hint}
H_{\rm int} = i \mu \sum_{j=1}^N \psi^L_j \psi^R_j
\eeq
 In the large $N$ limit,  $\mu$ and $J$ stay fixed.
We will mostly consider the case $\mu/J \ll 1$ and also consider temperatures $T/J \ll 1$. This will be true even for what we call ``high'' temperatures. 

  As an aside, let us mention that we can couple the two systems by an operator of dimension $\Delta$ 
\beq
\label{eq:hintoth}
H^\Delta_{\rm int} = i^F \mu (\Oc_\Delta)_L (\Oc_\Delta)_R
\eeq
where $F$ is the fermion number of $\Oc_\Delta$. We will mostly consider the case of $(\ref{eq:hint})$ which corresponds to $\Delta=1/4$ at low energies. However, we will
give certain estimates for generic $\Delta$. 

Like a single SYK, this model, \nref{CoupMod},  is solvable in the large $N$ limit. 
 We have four types of correlators: $G_{LL},G_{RR},G_{LR},G_{RL}$, each defined in Euclidean space as  
 \beq
G_{ab}(\tau) = \bra T \psi_a(\tau) \psi_b(0) \ket ~,~~~~{\rm with }~~~ a, \, b = L, \, R
\eeq
Since we are dealing with Majorana fermions we have  
\beq
G_{LL}(0^+)=G_{RR}(0^+)=\oh
\eeq
We have a closed system of Dyson--Schwinger 
 equations 
for the two point functions \cite{MQ}
\begin{align}
\label{eq:ds}
\pr_\tau G_{ab}(\tau) = \sum_{c} ~ \Sigma_{ac} * G_{cb}(\tau) + \delta_{ab} \delta(\tau) \\
\Sigma_{ab}(\tau) = J^2 G_{ab}(\tau)^3 - i \mu_{ab} \delta(\tau) ~,~~~~~{\rm no~sum}
\end{align}
where the convolution $*$ is taken along the Euclidean circle, $a,b=L,R$, and $\mu_{ab}$ is given by
\beq
\mu_{ab}=
\begin{pmatrix}
0 & \mu \\
-\mu & 0 \\
\end{pmatrix}
\label{mu_ab}
\eeq
The system has a $\mathbb{Z}_4$ symmetry $ \psi_L \ra -\psi_R$, $ \psi_R \ra \psi_L
$. 
%
Throughout our numerical computation we do not assume that this symmetry in unbroken.
We find that it is unbroken, since the (Euclidean) correlators we obtained obey the following relations:
\be 
G_{LL} = G_{RR} \ \text{pure real} ~,~~~~~~~
G_{LR} = -G_{RL} \ \text{pure imaginary}
\ee 
A convenient expression for the energy is 
\beq
\frac{E}{N}= \frac{J^2}{4} \int_0^\beta d\tau\ \l -G_{LL}^4 - G_{RR}^4 - 2 G_{LR}^4 \r + i \mu G_{LR}(0)
\label{eucl_energy2}
\eeq

The interaction term (\ref{eq:hint}) is a relevant perturbation, since for a single SYK model  the 
fermion $\psi$ has dimension $\Delta = 1/4$. 
Therefore at relatively high temperatures we expect that we have two weakly coupled
SYK models, whereas at low temperatures the system flows into a gapped phase with a gap 
that scales as \cite{MQ}  
%
\beq
E_{\rm gap} \propto \mu^{2/3} J^{1/3} ~,~~~~~~~~{\rm for } ~~~{ \mu \over J } \ll 1,\ {\rm and} \ 
q=4,\ \Delta=\frac{1}{4}
\eeq
Moreover the ground state 
is  close to the TFD of the two models with effective (inverse) temperature $\tilde{\beta}$:
\beq
| \text{TFD} \ket =  \sum_n e^{-\tilde{\beta} E_n/2} | \bar{E}_n \ket_L \times | E_n \ket_R ~,~~~~~~ 
\tilde \beta J \propto  \left( { J \over \mu } \right)^{ 1 \over 2 (1-\Delta) } ~,~~~\Delta ={1 \over q	 } 
\eeq
The energy of the ground state, relative to the energy of the two decoupled SYK models, scales as
\beq
 E_{\rm G} - 2 E_{0, \rm SYK}  \propto - N \frac{\mu^{4/3}}{J^{1/3}}
\eeq
And for general $\Delta$, $ E_G - 2 E_{\rm 0,SYK}  \propto - \mu^{\frac{1}{1-\Delta}}$

Since there is a gap and the ground state is unique,  the entropy is small in the ``cold wormhole phase".  
 Whereas in the two black hole phase,  we have a big entropy $2 S_0$, where  
  $S_0 \approx N \times 0.23$ is the ``zero temperature'' entropy of a single SYK model. 
The transition temperature $T_c$  is estimated by 
\beq
T_c \sim - \frac{E_{\rm G}}{2 S_0}  \propto { \mu^{ 4 \over 3 } \over J^{1\over 3} } , ~~~~~~~~~~{\rm for } ~~q=4
\eeq

For arbitrary temperatures the Dyson-Schwinger equations can be solved numerically by starting 
from $G_{LR}=G_{RL}=0, G_{LL}=G_{RR}=\half $ and then using an 
iteration procedure similar to one described in \cite{MS}. After obtaining the solution for some value of temperature, we
can use it as a seed for the iteration procedure at higher/lower temperature.
Figure \ref{fig:energy_fancy} shows energy as a function of beta for particular values of parameters $J=0.5, \mu=0.0053$.

\subsection{Perturbation theory at high temperature}
\label{sec:pert}
Here we use the term ``high" temperatures for temperatures for the two black hole branch of
the diagram  $T \gg \T2BH $, but still $T/J \ll 1$.  

For $\mu =0$ we have two copies of the conventional SYK correlators \cite{SachdevYe,kitaev}
\begin{align}
G(\tau)_{LL}=G(\tau)_{RR} = b \sgn(\tau) \frac{1}{\sqrt{J \beta}} \l \frac{\pi^2}{\sin^2  \frac{\pi \tau}{\beta}} \r^{1/4}, 
  ~~~~~~~G_{LR} =0 
\label{gll_conf}
\end{align}
with $b^{4} =1/( 4 \pi)$. 

 Now we turn on a small value of $\mu$ \nref{eq:hint}. 
If we are at sufficiently high temperature then the coupled system is still in the
phase with two separate black holes \cite{MQ}.
Nonetheless, the correlation between left- and right- SYK is not zero.
We can try to use the conformal perturbation theory to study the system.

\begin{figure}[!ht]
\minipage{0.47\textwidth}
\includegraphics[scale=0.9]{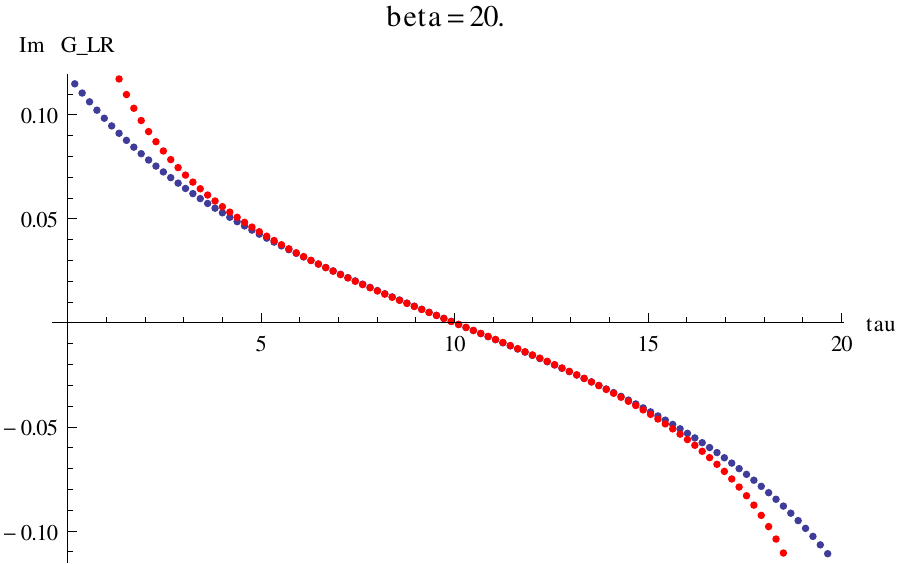}
\label{glr_comparison20}
\endminipage
\minipage{0.47\textwidth}
\includegraphics[scale=0.3]{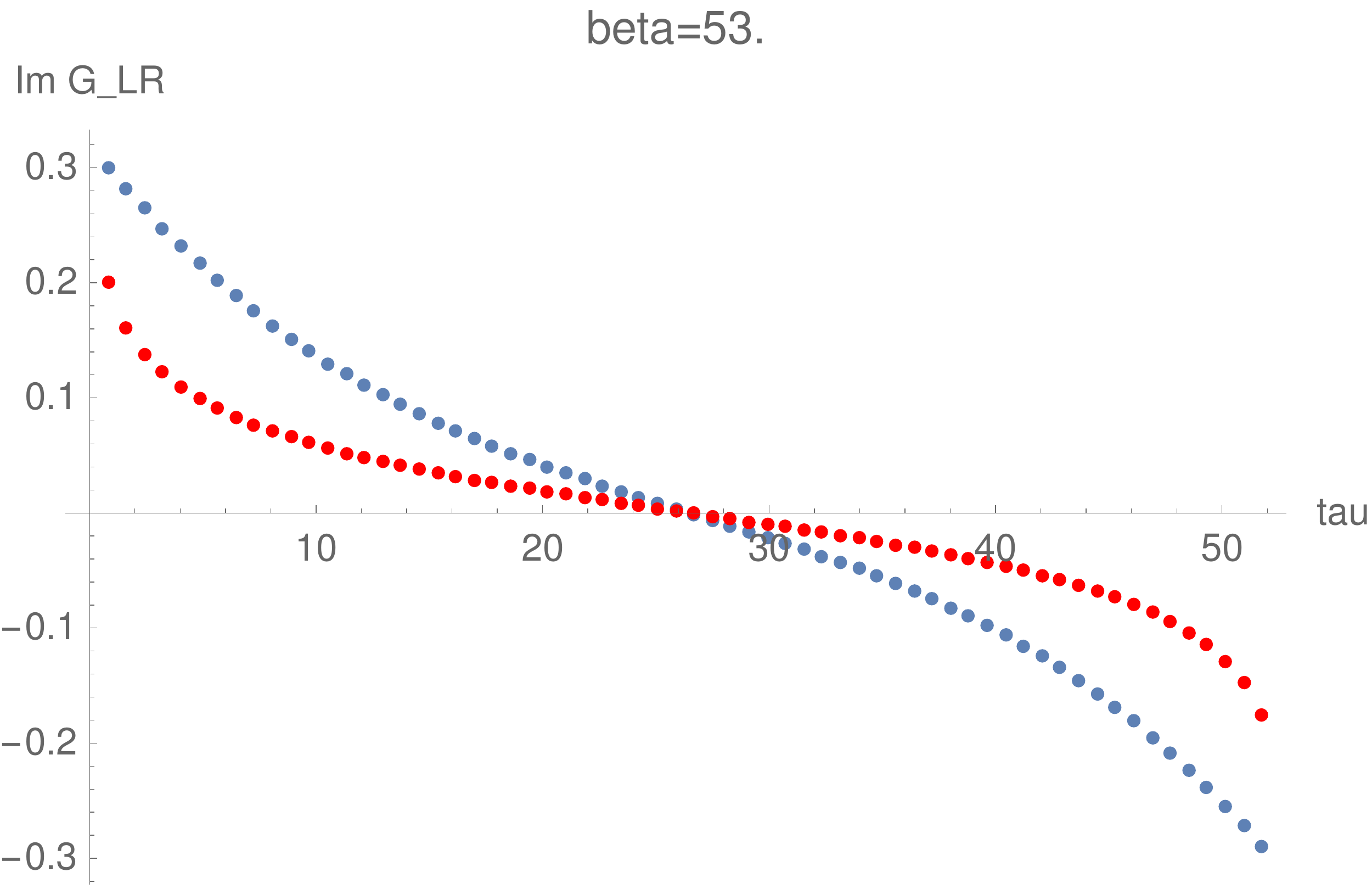}
\label{glr_comparison60}
\endminipage \\
\caption{\small Euclidean Green function $G_{LR}$.  The blue points correspond to the  exact solution, and the red ones  to the conformal answer (\ref{glr_conf}). Left: $\beta=20$. Right: $\beta=53$. 
For this values of parameters the transition to the wormhole phase happens around $\bet2BH \sim 61$. The approximation is better for higher temperatures. }
\label{glr_comparison}
\end{figure}

To linear order in $\mu$, only $G_{LR}$ Green functions receive a correction:
\beq
\Delta G_{LR}(\tau_1-\tau_2) =  i \mu \int_0^\beta d\tau \ G_{LL}(\tau_1-\tau) G_{RR}(\tau-\tau_2) 
\label{glr_int}
\eeq
This integral is computed analytically in Appendix \ref{app:delta_glr}. 
We can compare this leading order approximation against exact numerical solution of the Dyson--Schwinger equation for $J=0.5, \mu=0.05$ and different $\beta$s, 
see Figure \ref{glr_comparison}.

Also in Appendix \ref{app:delta_glr} we computed the $\mu^2$ correction to the energy. So that at low temperatures
we have the following expression for the energy:
\bea  
\label{delta_E}
E&=&2E_\text{SYK} + \Delta E \\
 E_\text{SYK}& =& E_\text{0,SYK} + \frac{2 \pi^2 \alpha_S \sqrt{2}}{J \beta^2} \la{EnSYK} \\
 E_{\rm 0,SYK} &=& -J \times 0.04063(1), ~~~~~~~~~~~\alpha_S=0.0071 \pm 0.0004 \la{AlphaS} \\
\Delta E &=& \frac{1}{2\sqrt{4 \pi}} \frac{\mu^2}{J} \l 2 \log \frac{\pi}{\beta J} -2 - c_1 \r \nonumber, \  ~~~~~~~~~~~~c_1\ =1.66(1)
\eea
where $E_\text{SYK}$ is the low-temperature result for a single SYK \cite{MS} and $\Delta E$ is the leading $\mu^2$
correction derived in Appendix \ref{app:delta_glr}. 
The comparison between \nref{delta_E} and the numerics is presented on Figure \ref{sch_comp}(b).

This approximation works better for high temperatures and then deteriorates close to  $\T2BH$, where the phase is supposed to connect with the hot wormhole phase. 


 Let us find out until what temperature we can trust the perturbation series in $\mu$, \nref{eq:hint}. 
The first point is that only even terms are non-zero. The term of order $\mu^{2n}$ contains
  a  $2n$-point function on the left and $2n$-point function on the right, each of these now computed in a single SYK model. We are only interested in connected correlators for computing the corrections to the free energy. 
Higher point functions in SYK have two contributions: a purely conformal piece which is independent of $\beta$, up to an overall factor of  
$\l 1/\sqrt{J \beta} \r^{4n}$ as in \nref{gll_conf},   plus contributions 
from the Schwarzian which are enhanced by an extra factor of 
$\beta J$. We claim that the Schwarzian contributions are in fact zero, see Appendix \ref{sch:1loop}. 
 The reason behind this cancellation is the following. When we work at large $N$ we are solving the classical equations. The reparametrization mode has a solution that is time translation invariant. The two sides are coupled by convolutions of Green's functions but  this translation symmetry remains unbroken. This means that there is no source for higher Fourier components of the reparametrization mode, so that the standard thermal solution continues to be a solution. 

The integrals over time give  $\beta^{2n}$. In total, we  
have $\frac{\mu^{2n} \beta^{2n}}{(J \beta)^n}$. 
So the expansion parameter is 
\be
\la{ExpPar} { \mu \beta \over \sqrt{J \beta} } ~,~~~~~{\rm for } ~~q=4 ~,~~~~{\rm or }~~~~~ { \mu \beta \over ( J \beta )^{ 2 \Delta }} ~,~~~~~~~{ \rm for ~general~}q
\ee
So we can trust the above perturbative answer until temperature 
\beq
T_{\rm pert} \propto J \left( { \mu \over J } \right)^{\frac{1}{1-2 \Delta}} 
\eeq
For $\Delta=1/4$ this scales as $\mu^2$, whereas   $T_c \propto  \mu^{4/3}$ is much larger. 

\begin{figure}[!h]
\centering
\includegraphics{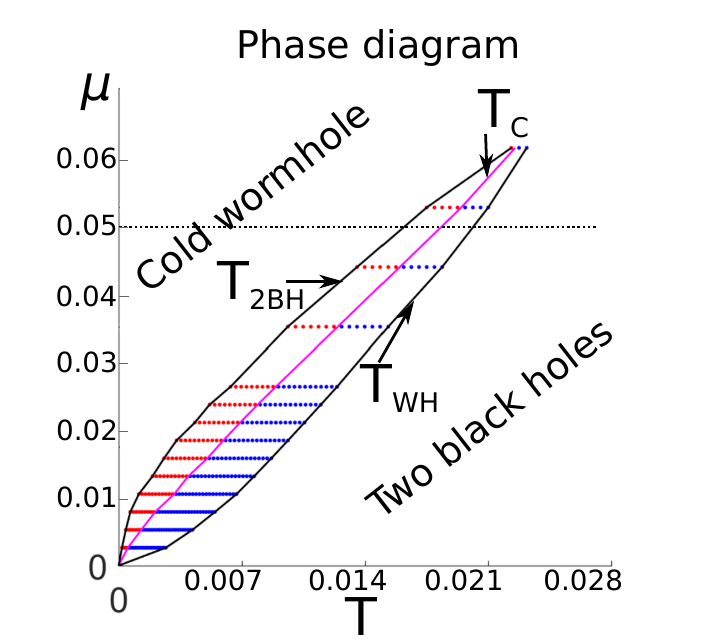}
\minipage{0.8\textwidth}
\caption{\small A reprint of the phase diagram obtained numerically in \cite{MQ} for $J=0.5$.
The right solid black curve indicates $\TWH$, purple line $T_c$ and  left solid black line $\T2BH$.
The dashed horizontal line is at $\mu=0.05$ ,  the value of $\mu$ we will use in our real time numerical simulation.
In this case $\bet2BH \sim 61, ~\beta_c \sim 54,~ \betWH \sim 49$.  }
\label{mq_phase}
\endminipage
\end{figure}

\begin{figure}[!h]
\centering
\minipage{0.49\textwidth}
\includegraphics[scale=0.5]{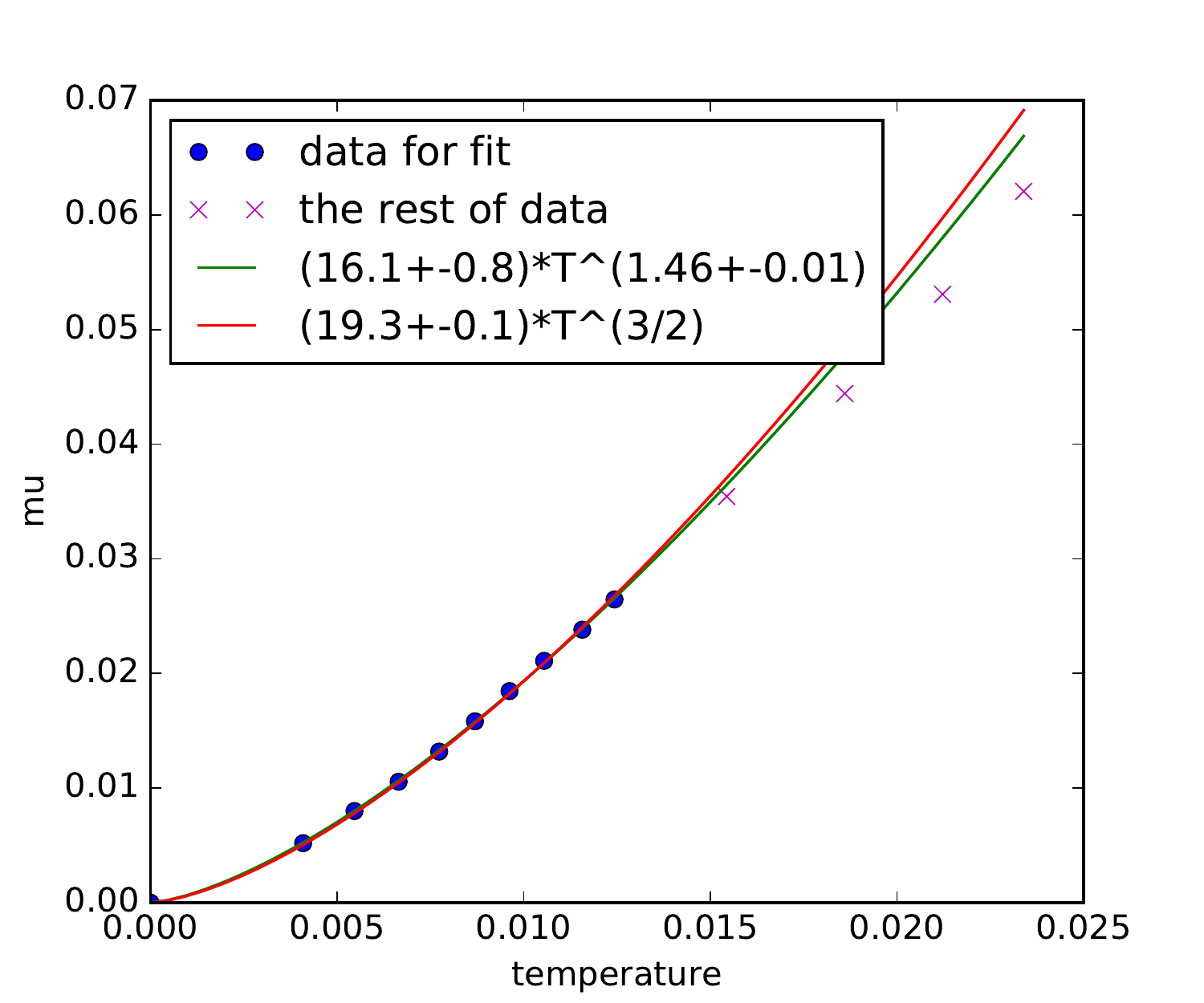}
\caption{\small The fit for $\TWH$ using the numerical data from \cite{MQ} in Figure \ref{mq_phase}, 
using only   data points with $\mu < 0.03$. The fit is consistent
with the analytical prediction $\mu \sim T^{3/2}$.}
\endminipage\hfill
\minipage{0.49\textwidth}
\includegraphics[scale=0.5]{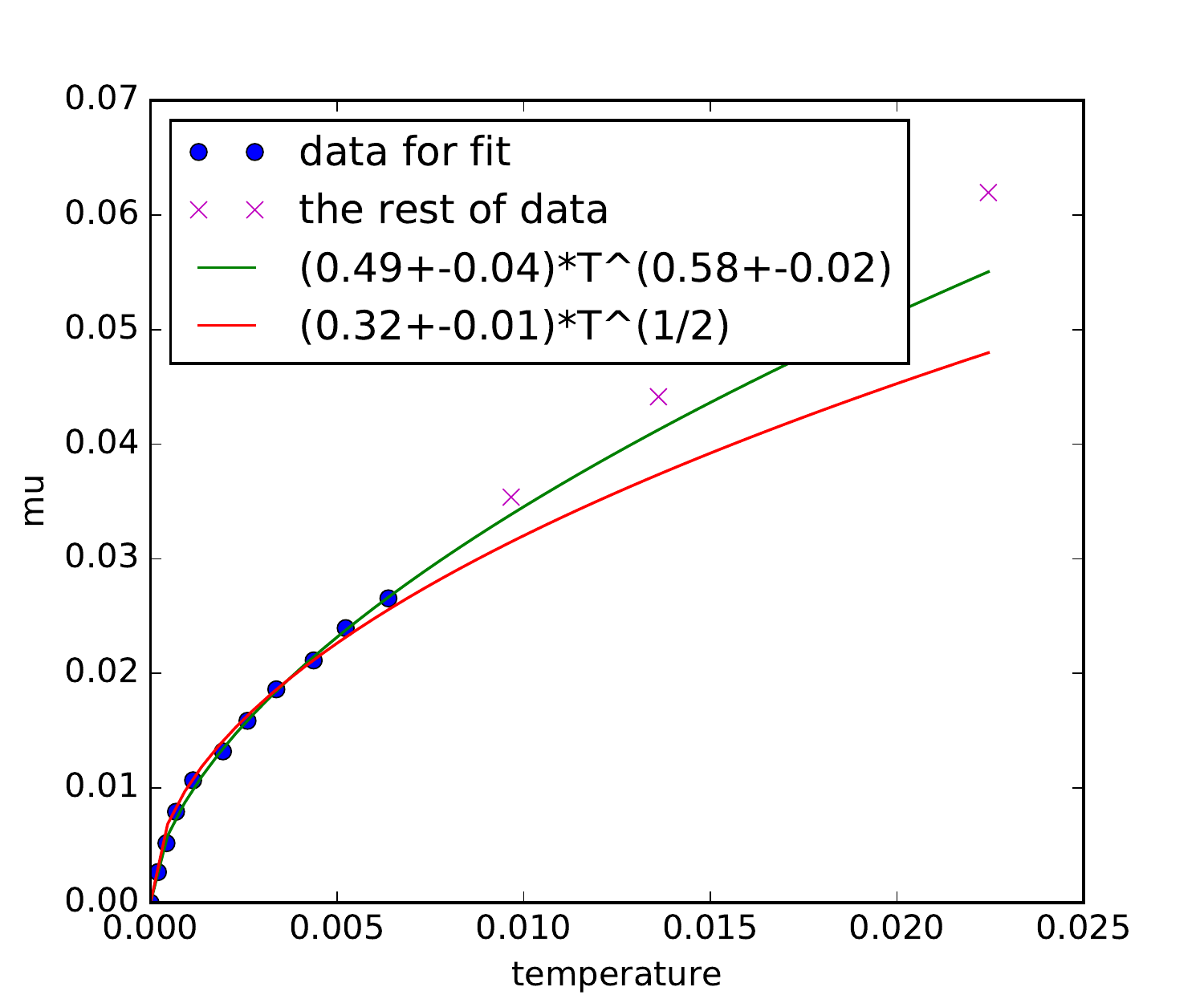}
\caption{\small The fit for $\T2BH$ using the numerical data from \cite{MQ} in Figure \ref{mq_phase}, 
using only  data points with $\mu < 0.03$. The fit is consistent
with the analytical prediction $\mu \sim \sqrt{T}$.}
\label{mt_fit}
\endminipage\hfill
\end{figure}

We conjecture that the transition temperature $\T2BH$, when the two black holes phase cease to exist,   in fact coincides
with $T_{\rm pert}$, when the perturbation theory in $\mu$ breaks down
\beq
\label{T1_est}
\T2BH \sim T_{\rm pert} \propto  J \left( { \mu \over J } \right)^{\frac{1}{1-2 \Delta}}
\eeq
We check this  prediction for $\Delta=1/4$ against the numerical phase diagram obtained 
in \cite{MQ}. See Figures   \ref{mq_phase}, \ref{mt_fit}. 

One last comment on the leading result (\ref{glr_int}). In Appendix \ref{sch:1loop}, we studied
the gravitational dressing of this term, searched for instabilities that would spontaneously break the $U(1)$ time translation symmetry,  but did not find any.    

\subsection{Low temperature thermodynamics using the Schwarzian }
\label{sec:sch_low}

In this section we review the results of \cite{MQ} on the Schwarzian description of the wormhole. We will see that the Schwarzian indeed admits a wormhole
solution at low temperatures. Moreover, by including the matter contribution to the partition function one is able to see two phases which join at
(inverse) temperature $\betWH$. One phase has positive heat capacity and almost zero entropy. This is a cold and mostly 
empty wormhole. The second phase 
has higher energy longer throat and negative heat capacity. This is wormhole with extra matter excitations inside. We will see shortly that at small
enough $\mu$ this approximation predicts $\betWH$ with good accuracy. 

As we have mentioned before, the ground state of the system is close to the TFD state. Since we have global $AdS_2$ we have the following
$G_{LR}$ correlator in Euclidean signature:
\beq
G_{LR} = \frac{i b}{\sqrt{2 J \cosh \l \frac{t_L- t_R}{2} \r}}
\eeq
where $t_L, t_R$ are times on left/right.

The action in the Schwarzian approximation now includes two kinetic terms\footnote{For $q=4$, $J=\sqrt{2} \Jc$ hence the extra $\sqrt{2}$.} and an interaction
\bea
S &=& S_{\rm kin} + S_{\rm int} 
\\
S_{\rm kin} &=& - \frac{N \sqrt{2} \al_S}{J} \int du \ \l \{ \tanh \l \frac{t_L(u)}{2} \r , u  \} 
+ \{ \tanh \l \frac{t_R(u)}{2} \r , u  \} \r 
\\ 
S_{\rm int} &=& -  N i  \mu \int du G_{LR}(u) = \frac{N \mu b}{\sqrt{2 J}} \int du \frac{( \tpr_L(u) \tpr_R(u) )^{1/4}}{\sqrt{\cosh \l \frac{{\bf t}_L(u) - {\bf t}_R(u)}{2} \r}}
\eea
with $\alpha_S$ in \nref{AlphaS}. 
The wormhole solution is simply ${\bf t}_L = {\bf t}_R = \tpr u$, where $\tpr$ is a constant. The effective temperature, $\tilde{\beta}$ of the  TFD state is 
given by 
\beq
\tilde{\beta} = \tpr \beta
\eeq
Inserting into the action we get  the free energy
\beq
 F/N = \frac{\al_S \sqrt{2}}{J} \tpr^2 - \frac{\mu b}{\sqrt{2J}} \tpr^{1/2}
\eeq
We should also include the contribution from matter fields in the wormhole throat. If the temperature is low enough we excite only the 
lightest excitation in the bulk, which is the elementary fermion with mass $1/4$.
Its contribution to the free energy is:
\beq
\Delta  F/N = - \frac{1}{\beta} \log \l 1 + e^{-\tilde{\beta}/4} \r \approx - \frac{1}{\beta} e^{-\tpr \beta/4}
\eeq
Extremizing the free energy with respect to $\tpr$, which is the same as extremizing the 
full effective action,  we have the following equation to determine $\tpr$ and correspondingly $\tilde{\beta}$:
\beq
\label{saddle}
0=\frac{\pr F_{\rm tot}}{\pr \tpr} = \frac{2\al_S \sqrt{2}}{J} \tpr - \frac{\mu b}{2\sqrt{2J}} \tpr^{-1/2} + \frac{1}{4} e^{-\tpr \beta/4} 
\eeq

\begin{figure}[!ht]
\begin{minipage}{0.47\textwidth}
\includegraphics[scale=0.35]{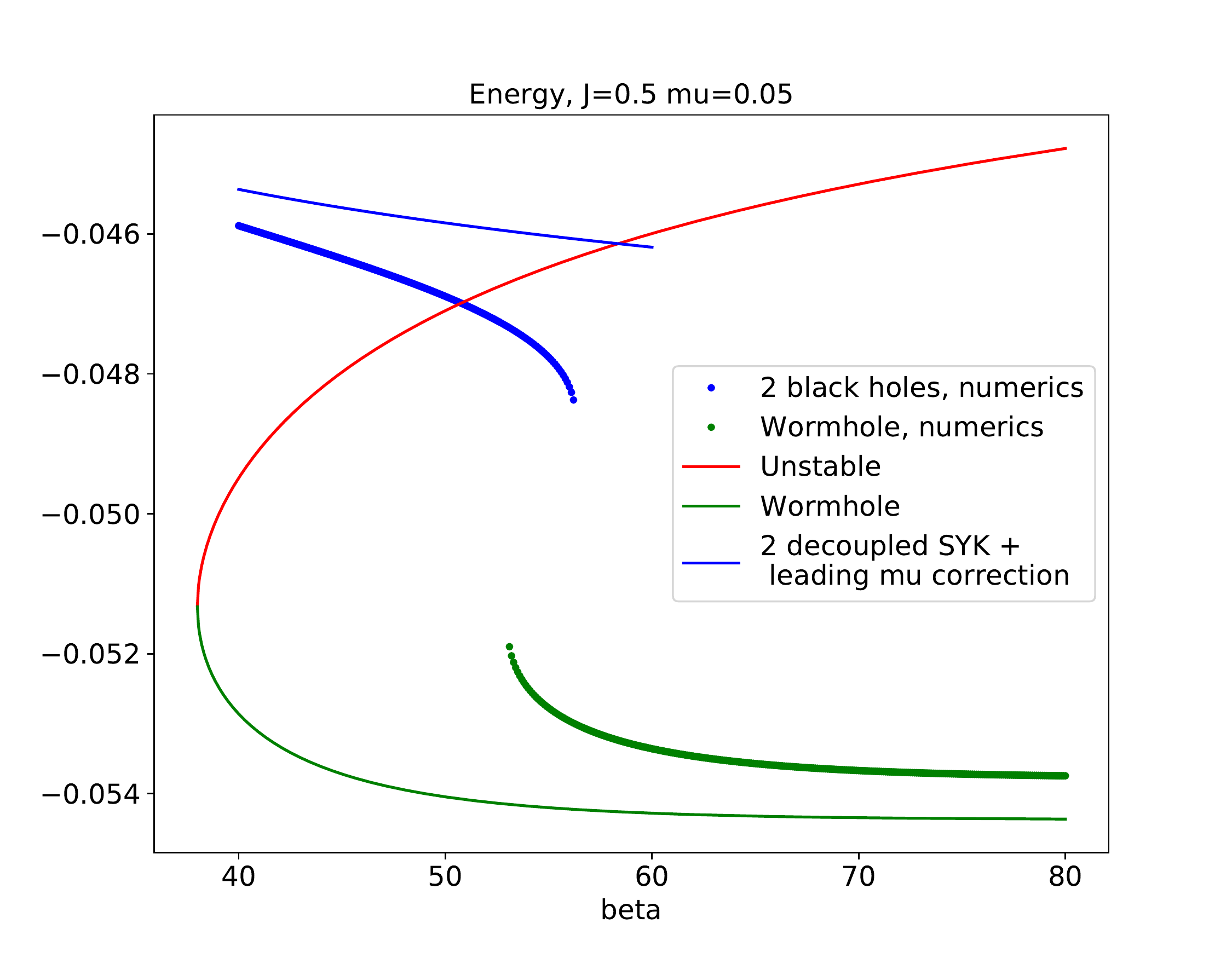}
\end{minipage}
\begin{minipage}{0.47\textwidth}
\includegraphics[scale=0.35]{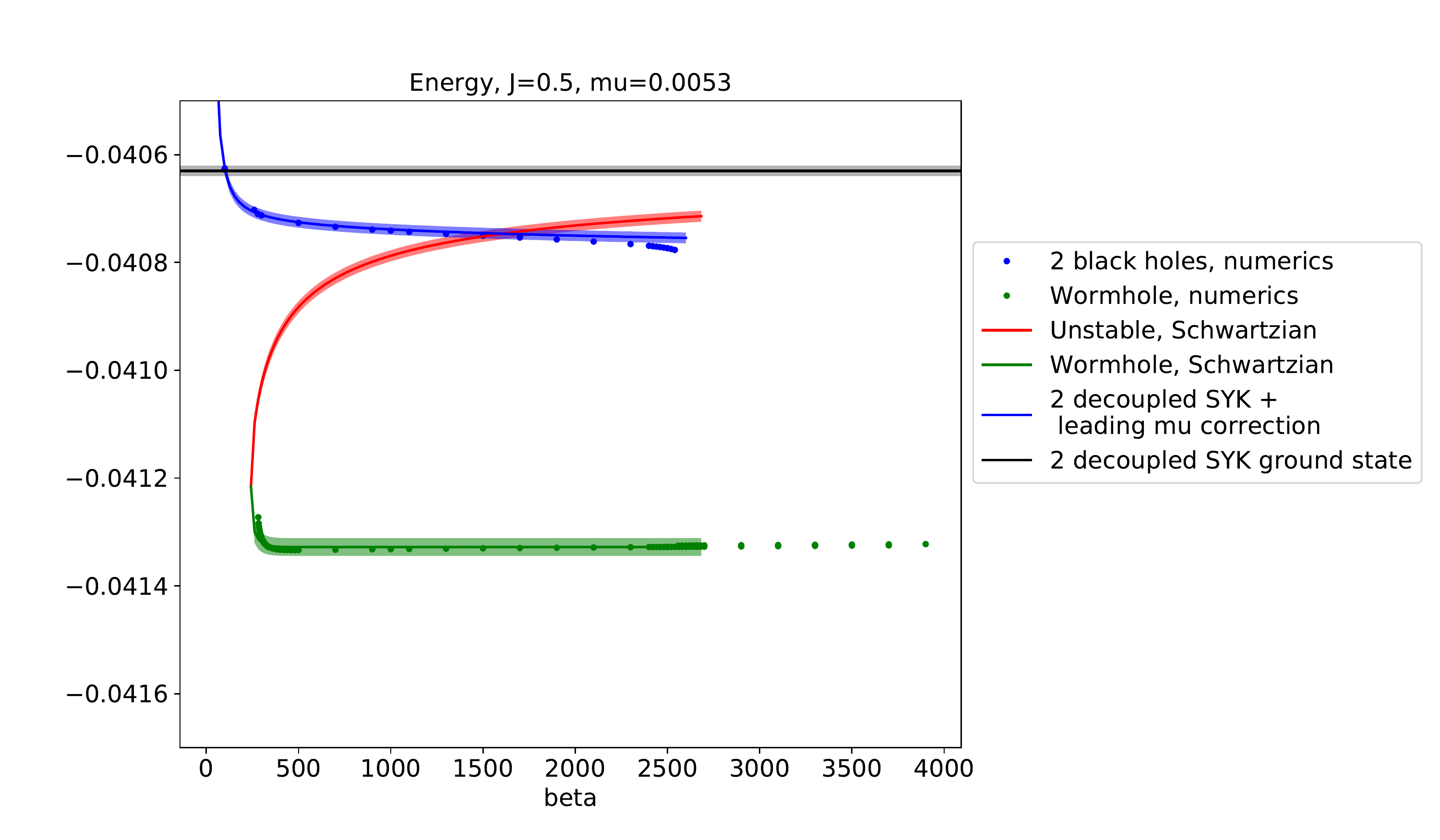}
\end{minipage}
\begin{center}(a) ~~~~~~~~~~~~~~~~~~~~~~~~~~~~~~~~~~~~~~~~~~~~~~~~~~~~~~~~~~~~~~~~~(b) 
\minipage{0.8\textwidth}
\caption{\small Thermodynamics of the model for two different values of $\mu/J$. In 
both $J=0.5$ and in (a) $\mu =0.05$ and in (b) $\mu = 0.0053$. The dots correspond to numerical solutions of the DS equation. The blue ones for the two black hole phase and the green ones for the cold wormhole phase. 
The black line is the ground state energy of two decoupled  SYK, $2 E_{0,SYK}$ (on the left plot it lies above the visible area). 
Blue curve: the perturbative result (\ref{delta_E}) for the energy.
Green curve: wormhole branch of eq. (\ref{saddle}). 
Red curve: unstable branch of eq. (\ref{saddle}). 
The uncertainties, represented as shaded regions, come from the uncertainties in $E_{\rm 0,SYK}$ and $\alpha_S$. We see that in (b) the agreement is very good. However, in (a) the agreement is not so good, but the qualitative form of the curve is similar, if we joint the two end points of the dotted lines. }
\label{sch_comp}
\endminipage
\end{center}	
\end{figure}

This equation  has a solution with very small $\tpr$ which we can not trust, because we were assuming 
that the temperatures are low. 
For large enough $\beta$ two additional solutions emerge. One of them correspond to what we call the cold wormhole and the other to the hot wormhole. The latter has negative heat capacity and can be viewed as a wormhole with more excitations in the throat.  
Figure \ref{sch_comp} shows the two branches for two different values of $\mu/J$ and their comparison to the numerical solution to the DS equations.

We can calculate $\TWH$ from this equation.  We simply need to find when these two solutions merge. 
To keep the discussion general, we consider general $\Delta$, which corresponds to the case when
the two sides are coupled through the product of two operators of dimension $\Delta$. The equation for $\tpr$ now reads as:
\beq
\label{Constr}
\frac{2\al_S \sqrt{2}}{J} \tpr - \frac{\mu b_\Delta}{2 (2J)^{2 \Delta}} \tpr^{2 \Delta-1} +  \Delta e^{-\tpr \beta \Delta} = 0
\eeq
The cold wormhole branch can be approximately found \cite{MQ} by neglecting the thermal excitations exponent in the above equation, so that $\tpr$ is equal
\beq
\label{t_prime_wh}
\tpr \propto \l \frac{\mu}{J^{2 \Delta-1}} \r^{\frac{1}{2- 2 \Delta}}
\eeq
Whereas the unstable branch with excitations can be approximated by neglecting the Schwarzian kinetic term $\sim \al_S \tpr$:
\beq
\label{UnstT}
\frac{\mu}{J^{2 \Delta}} \tpr^{2 \Delta-1} \propto e^{-\tpr \beta \Delta}
\eeq
Plugging the $\tpr$ from the first solution (\ref{t_prime_wh}) into the above equation we find the $\TWH$:
\beq
\label{eq:T_l}
\TWH \propto \frac{\mu^{\frac{1}{2-2 \Delta}}}{\log J/\mu} \sim \frac{\mu^{2/3}}{\log J/\mu}, \ \text{for}\ \Delta=1/4
\eeq
Of course, both \nref{t_prime_wh} and \nref{UnstT} are good for 
$T\ll \TWH$. Here we presented just estimates, but it is straightforward to solve \nref{Constr} numerically, see Figure 
\ref{sch_comp}. 

We can also compute the energy using\footnote{When computing the derivatives one has to keep in mind that $\tpr$ is a function of $\beta$.  And use  (\ref{saddle}).
} 
\beq
\label{saddle_energy}
E = F + T S = F + \beta \frac{\pr F}{\pr \beta} + \beta \frac{\pr F}{\pr \tpr} \frac{d \tpr}{d \beta} =2 E_{\rm 0,SYK} +  \frac{\al_S \sqrt{2}}{J} \tpr^2 - \frac{\mu b}{\sqrt{2J}} \tpr^{1/2} + \frac{\tpr}{4} e^{-\tpr \beta/4}
\eeq
where $E_{\rm 0,SYK}$ is the ground state energy of a single SYK, see \nref{AlphaS}. 

In Figure \ref{sch_comp} we have compared the results computed 
using eq. (\ref{saddle_energy}) with the numerical solution of the DS equation. 
For the value $\mu=0.05$, which is the one we will use for the real time numerical computation, the agreement is not very good, but the qualitative form of the curve is similar, see Figure \ref{sch_comp}(a). 
  This means that that $\mu$ is not low enough for an accurate   Schwarzian description. Indeed if we lower $\mu$ we get very good agreement. See Figure \ref{sch_comp}(b) for   $\mu = 0.0053$. 

To summarize, for small $\mu$, we have a hierarchy of  temperatures
\begin{align}
\TWH & \propto \frac{\mu^{\frac{1}{2-2\Delta}}}{\log J/\mu} \sim ~\frac{\mu^{2/3}}{\log J/\mu}, \ ~~~~\text{for}\ \Delta=1/4 \nonumber \\
T_c & \propto \mu^{\frac{1}{1-\Delta}} \sim \mu^{4/3} \\
\T2BH & \propto \mu^{\frac{1}{1-2 \Delta}} \sim \mu^{2} \nonumber
\end{align}
where the rightmost term corresponds to $\Delta =1/4$. 

\section{Real time results}
\la{RealTime}

\subsection{Coupling to a bath}

\label{sec:bath}
In order to study the real-time formation of the wormhole, we need to cool down the system. 
Before considering real-time dynamics, first we need to understand how to couple 
our system to a thermal bath. Ideally we want the bath to be a large system in order to avoid back-reaction. 

Generally, we can couple a system's operator $\Oc_S$ to a bath operator $\Oc_B$:
\beq
\Delta S = i^F V \int d\tau \  \Oc_S \Oc_B
\eeq
where $F$ is the fermionic number of $\Oc_S$.

If $V$ is small and the bath is large we can study this interaction in the Lindblandian approximation by
considering the 1-loop result and assuming that there is no back reaction on the bath, such that
we can substitute the product $\Oc_B(t_1) \Oc_B(t_2)$ by VEV $\bra \Oc_B(t_1) \Oc_B(t_2) \ket_B$:
\beq
\Delta S = - V^2 \int d\tau_1 d\tau_2 \ \Oc_S(\tau_1) \Oc_S(\tau_2) \bra \Oc_B(\tau_1) \Oc_B(\tau_2) \ket_B 
\eeq

For our problem we have a varying temperature that sets an energy scale for the model. Specially for our numerical analysis, it is convenient to choose an interaction that is  scale invariant (at least approximately), so that the effects of coupling to the bath are independent of the temperature. Otherwise the bath might be effectively decoupling in some temperature range and the system would take long to cool down. 


A natural model for the bath is 
 another SYK, possibly with larger number of fermions.
We can consider the bath to be another single-SYK model with $q=q_B$ with large number of fermions $M$, 
much larger than the number of fermions in our system $N$.
Recently this problem was studied in \cite{syk_bath}, we refer to this paper for details.  If we
denote the bath fermions by $\chi_\al$, then the coupling can involve an arbitrary number $s$ of system fermions and $p$ bath fermions. 

Again, to warm-up, let us first consider the case when the system consists of a single-SYK model. The coupling between the bath and the system has
the form:
\beq
\Lc_\text{bath-system} = V_{i_1 \dots i_s}^{\al_1 \dots \al_p} \psi_{i_1} \dots \psi_{i_s} \chi_{\al_1} \dots \chi_{\al_p} 
\eeq
where $V$ is a random Gaussian-distibuted coupling. 
By choosing its variance appropriatly, the back reaction can be made of order $N/M \ll 1$. 
The Euclidean Dyson--Schwinger equations for the system
stay the same except for a correction to the self-energy: the self-energy acquires an additional term,
\beq
\Delta \Sigma = \Sigma_B = \eta G_{B}^p G^{s-1}
\eeq
where $G_B$ is bath two-point function and $\eta$ is proportional to the variance of $V$. 

We can get a marginal interaction with 
 $s=1, p=3$,    when the bath consists of $q=4$ SYK models 
 (with, say    $J_B=J_S$). 
This is the bath we will use.
%
%
More precisely, we introduce two separate baths, one for each SYK factor
\beq
\Lc_\text{bath-system} = V_{i}^{\al_1 \al_2 \al_3} \psi_{L,i} \chi_{\al_1} \chi_{\al_2} \chi_{\al_3} + 
\tilde{V}_{i}^{\al_1 \al_2 \al_3} \psi_{R,i} \tilde{\chi}_{\al_1} \tilde{\chi}_{\al_2} \tilde{\chi}_{\al_3} 
\eeq
where $V$ and $\tilde{V}$ are independent Gaussian-distributed variables. This interaction leaves $\Sigma_{LR}$ unperturbed, but the other two
self-energies have additional terms now:
\beq
\Delta{\Sigma}_{LL} = \Delta{\Sigma}_{RR} = \Sigma_B = \eta G_B^3
\eeq
The above equations are written in  Euclidean signature. 
We now turn to Lorentzian equations.

\subsection{Kadanoff--Baym equations}
\label{sec:kb_intro}

We now  write down the Lorentzian  time version of the DS equations. For details see Appendix \ref{ds:lorentz}.

 Here we will discuss the non-equilibrum situation following \cite{MS} and \cite{quench1}.
It is convenient to work right away on the Keldysh time contour, see   \cite{kamenev_book} for a comprehensive introduction. 
We will need Lorentzian time correlators which are not time ordered. This
can be achieved by introducing a Keldysh time contour which runs from $-\infty$ to $+\infty$ and then back from $+\infty$ to $+\infty$.
First, one introduces a Wightman function
\beq
G^>_{ab}(t^-_1,t^+_2) = 
-i \bra \psi_a(t^-_1) \psi_b(t^+_2) \ket,
\eeq
where $t_1,t_2$ lie on different sides of the contour. This simply means that $\psi_b(t_2)$ is always located to the right of $\psi_a(t_1)$
in the correlator, regardless of time-ordering. This is why the Keldysh contour consists of two parts.
Note the overall $i$ in front of the correlator. 
Since we are dealing with simple Majorana fermions the ``lesser'' function $G^<_{ab}$ is given by:
\beq
G^<_{ab}(t_1,t_2) = - G^>_{ba}(t_2,t_1)
\eeq
Also we will need retarded and advanced Green functions:
\begin{align}
G^R_{ab}(t_1,t_2) = \theta(t_1-t_2) \l G^>_{ab}(t_1,t_2) - G^<_{ab}(t_1,t_2) \r \\ \nonumber
G^A_{ab}(t_1,t_2) = \theta(t_2-t_1) \l G^<_{ab}(t_1,t_2) - G^>_{ab}(t_1,t_2) \r
\label{}
\end{align}
Dyson--Schwinger equations written on the Keldysh contour 
are known as Kadanoff--Baym equations, and are useful for non-equilibrium situations.   Let us write them down explicitly for a single SYK:
\begin{align}
i \pr_{t_1} G^>(t_1,t_2) = \Sigma * G =  \int_{-\infty}^{+\infty} dt \ \l \Sigma^R(t_1,t) G^>(t,t_2) + \Sigma^>(t_1,t) G^A(t,t_2) \r \nonumber \\
-i \pr_{t_2} G^>(t_1,t_2) = G * \Sigma =  \int_{-\infty}^{+\infty} dt \ \l G^R(t_1,t) \Sigma^>(t,t_2) + G^>(t_1,t) \Sigma^A(t,t_2) \r
\label{eq:kb_syk1}
\end{align}
 where the self-energy $\Sigma^>$ is given by 
\beq
\Sigma^>(t_1,t_2) = -J^2 \l G^>(t_1,t_2) \r^3
\eeq
These equations for the complete system of two interacting SYKs and a bath are derived using the path integral in Appendix \ref{eff_action_der}.

Remember that the ``greater'' Green function $G^>(t_1^-,t_2^+)$ has time arguments lying on different sides of the Keldysh contour, this is why we do not have
a delta-function on the right hand side of \nref{eq:kb_syk1}.
The integral in the right hand side of \nref{eq:kb_syk1}, which  involves different 
Green functions, is just a simple convolution $\Sigma * G$ along the Keldysh contour \cite{stefanucci2013}.
 We can show it by  writing the anti-time ordered $\Sigma = \Sigma^> - \Sigma^R$ and the time ordered 
$G = G^A + G^>$.
One can easily see that equations  \nref{eq:kb_syk1}  are casual.

Let us mention one subtlety. Strictly speaking, if one starts from a thermal state, then the precise Keldysh contour involves imaginary time stip at the end of the
lower branch, at $t=0$. This time strip has length $\beta$ and prepares the thermal state. One can bypass this as follows. 
First we set the coupling to the bath to zero. Then we find the real time correlators at thermal equilibrium by solving the 
 real time equations imposing the appropriate relations between the Green's functions, see e.g. \nref{GpGR}. We then evolve 
the system for some time $t\gg \beta$ and then we turn on the coupling to the bath. 
For more details see Appendix  \ref{numerics}.

For two coupled SYK models one has to be very careful with the $\mu$ term. 
To understand its form on the Keldysh contour we can go back to $G,\Sigma$ effective action derived in Appendix \ref{eff_action_der}
\begin{align}
i \frac{S}{N}=\frac{1}{2} \Tr \log \l \omega - \Sigma_{ab}(\omega) \r - 
\sum_{ab} \int_\Cc dt_1 dt_2 \l  \frac{J^2}{8}  G_{ab}(t_1,t_2)^4  + \frac{1}{2} \Sigma_{ab}(t_1,t_2) G_{ab}(t_1,t_2) \r + \nonumber \\
+\frac{i \mu}{2} \int_\Cc dt \ \l G_{LR}(t,t) -G_{RL}(t,t) \r  - \frac{\eta}{2} \int_{\Cc} dt_1 dt_2 \l G_{LL}(t_1,t_2) + G_{RR}(t_1,t_2) \r G_B^3(t_1,t_2)
\label{eq:eff_action}
\end{align}
We see that the $\mu$ contribution to $\Sigma(t_1,t_2)$ is
\beq
\Delta \Sigma_{ab}(t_1,t_2) = i \mu_{ab} \delta_\Cc(t_1-t_2)
\eeq
In this expression $t_1,t_2$ can be on either side of the Keldysh contour, 
this is why $\Sigma$ does not have an additional index, like $>,R,A$. 
Notice that the delta-function $\delta_\Cc$ is defined on the Keldysh contour as well. 
It yields non-zero answer if and only if $t_1=t_2$ and $t_1,t_2$  are on the same side of the contour.
Let us compute the contribitution of this term to the convolution $\Sigma * G$ along the Keldysh contour:
\beq
\Delta (\Sigma * G)(t_1^-,t_2^+)_{ab}  = i \mu_{ac} \int_\Cc dt \ \delta_\Cc(t_1^- - t) G_{cb}(t-t_2^+) = i \mu_{ac} G^>_{cb}(t_1^-,t_2^+)
\eeq

The final form of the Kadanoff--Baym equations, including the bath, is:
\begin{align}
i \pr_{t_1} G_{ab}^>(t_1,t_2) = i \mu_{ac} G^>_{cb}(t_1,t_2) + \int_{-\infty}^{+\infty} dt \ \l \Sigma^R_{ac}(t_1,t) G^>_{cb}(t,t_2) + \Sigma^>_{ac}(t_1,t) G^A_{cb}(t,t_2) \r  \nonumber \\
-i \pr_{t_2} G_{ab}^>(t_1,t_2) =  i \mu_{ac} G^>_{cb}(t_1,t_2) +  \int_{-\infty}^{+\infty} dt \ \l G^R_{ac}(t_1,t) \Sigma^>_{cb}(t,t_2) + G^>_{ac}(t_1,t) \Sigma^A_{cb}(t,t_2) \r
\label{eq:kb}
\end{align}
where $\mu_{ab}$ is defined in eq. (\ref{mu_ab}) and the self-energy is
\beq
\label{def:sigma}
\Sigma_{ab}^>(t_1,t_2) = - J^2 \l G_{ab}^>(t_1,t_2) \r^3 - \eta \delta_{ab} \l G_B^>(t_1,t_2) \r^3
\eeq

\subsection{Forming the wormhole}
\label{sec:formation}
Our numerical setup for solving KB equation (\ref{eq:kb}),(\ref{def:sigma}) is presented in Appendix \ref{numerics}.
We prepare initial Lorentz Green's functions using Lorentz-time Dyson--Schwinger equations described in Appendix \ref{ds:lorentz}.
Our initial Green's functions correspond to two interacting SYK models with non-zero $\mu$ at thermal equilibrium. 
In order to save   computational time,  the initial temperature is chosen close to (and slightly higher than)  the 
transition temperature $\T2BH$. 

We extract the temperature using the Fluctuation-Dissipation Theorem(FDT) for the two-point functions. Precisely, our numerical setup gives us the  $G^>$ and $G^<$ Green's functions.
At thermal equilibrium a certain combination of these Green's functions (eq. (\ref{tanh})) must be equal to $\tanh(\beta \omega/2)$ in the frequency domain. 
So we take the numerically obtained 
$G^{>}$ and $G^<$,  make a discrete Fourier transform and fit eq. (\ref{tanh}) with the tanh. We refer to Appendix \ref{numerics} for details about the precise choice of the
Green's functions and the frequency domain for fit.

Our benchmark parameters are $J=J_B=0.5$ (the system and the bath have the same coupling strength), $\mu=0.05$. 
In this case,  from Figure \ref{mq_phase} we expect $\bet2BH \sim 61,~ \beta_c \sim 54,~ \betWH \sim 49$.

\begin{figure}[!h]
\centering
\includegraphics[scale=0.7]{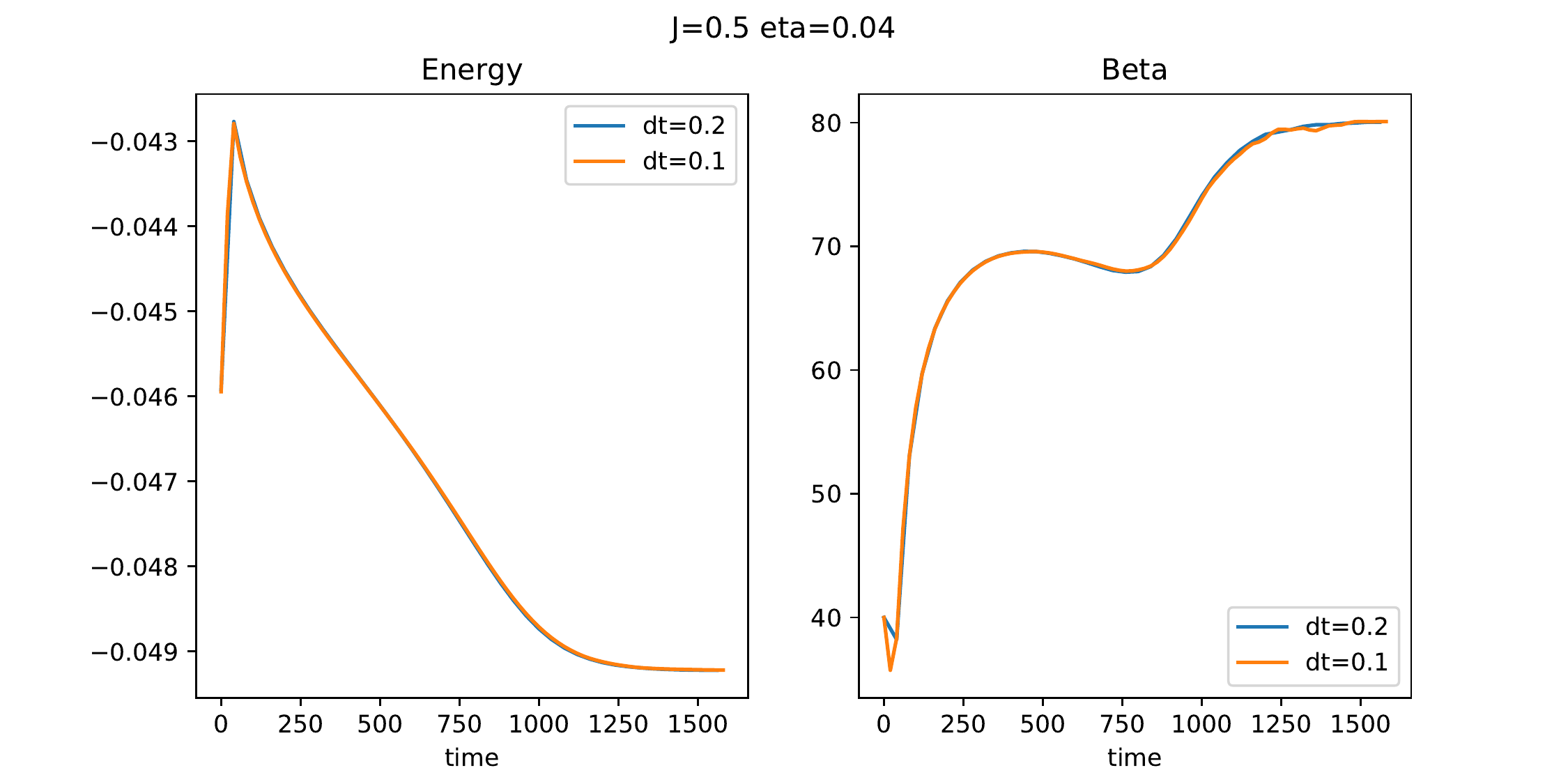}
~~~~~~~~~~~(a) ~~~~~~~~~~~~~~~~~~~~~~~~~~~~~~~~~~~~~~~~~~~~~~~~~~~~~(b) 
\minipage{0.8\textwidth}
\centering
\caption{ \small Numerical results for $J=0.5$ and $\mu = 0.05$, $\beta_B=80$ and 
timesteps $dt=0.2,0.1$. (a) The energy as a function of time. The initial rise is due 
to the fact that we are coupling the bath to the system, and this changes the energy \cite{syk_bath}.  
We then see the energy decreasing monotonicallly. (b) The inverse temperature as a 
function of time. We also see an initial sharp increase due to the coupling of 
the bath, then we see a decrease. Then a slight increase of the temperature that 
signals the phase with negative specific heat. Finally the temperature decreases again. 
}
\label{comparison_full}
\endminipage
\end{figure}

We considered a few values for the system-bath coupling, $\eta=0.04,~0.02,~0.01$. 
Figure \ref{comparison_full} shows the results for $\beta_{\rm bath}=80,~ \eta=0.04,~ \beta_{\rm initial}=40$ for different time steps. The energy is computed using eq. (\ref{kb_energy}).

\begin{figure}[!h]
\centering
\includegraphics[scale=0.5]{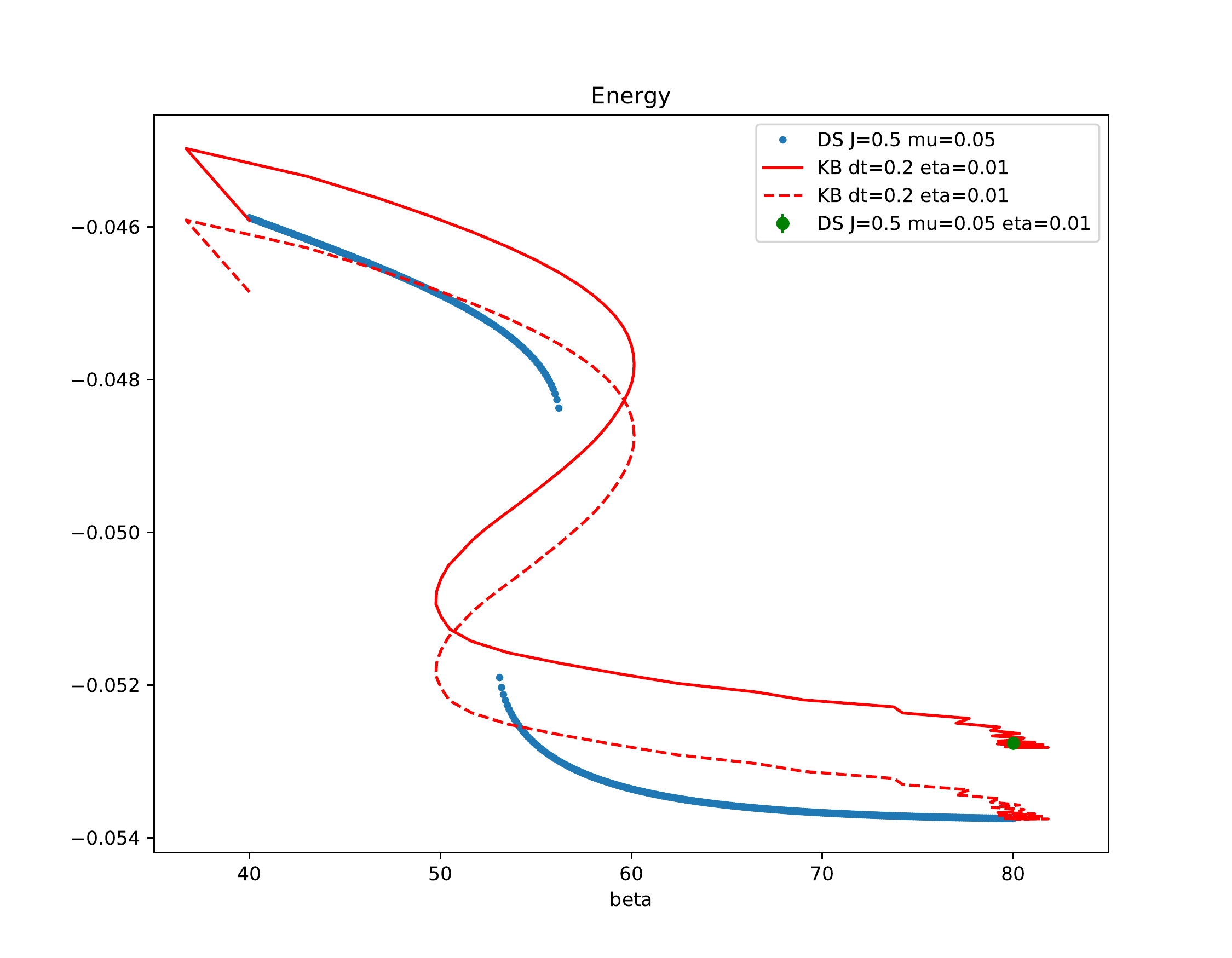}
\minipage{0.8\textwidth}
\caption{\small The comparison between $E(\beta)$ in equilibrium(blue points) and the real-time evolution of the system as it cools down (solid red line). The bath parameters are 
$\eta=0.01, ~\beta_{\rm bath}=80$. The initial linear ramp of the red curve occurs because we switch on the coupling with the bath linearly in time. 
The dashed red line is the same as the solid line but shifted to match the final energy on the phase diagram. 
The green dot shows the equilibrium value of the system+bath energy.}
\label{check001}
\endminipage
\end{figure}

We can perform another check, this time taking $\eta$ much smaller, namely $\eta=0.01$.
The result is shown on Figure \ref{check001}.
A few comments are in order. As is explained in detail in \cite{syk_bath} the marginal
system-bath interaction we will renormalize $J$, making it bigger. This is why expect that the actual transition will happen
at higher $\beta$. This is indeed what we see. Moreover,  the interaction with the bath will shift the ground state energy. To compensate for
this we have shifted the energy vs beta curve in Figure \ref{check001} to match the final energy.

The red curve in Figure \ref{check001} has wild oscillations in temperature near the equilibrium for small $\eta$, see also Figures \ref{comparison_full}(b). 
The reason  is the following.  Because of the numerical error there is an additional flux of energy which pushes the system out of equilibrium. 
From the phase diagram (blue points) we see that the derivative $d\beta/dE$ is very large. If $\eta$ is not big enough, the relaxation time is not small enough to smooth out
these fluctuations.

The crucial question is whether we indeed have reached \textit{the} wormhole phase or not. In principle, we might have ended up in some other phase.
To verify that we have reached the wormhole we can make a precise check of the system's energy.

In the real-simulation the final value of the energy is(in units where $J=0.5$):
\beq
E_{\rm KB} = - 0.05282(2)
\label{eta0012_energy}
\eeq
where the uncertainty comes from changing the size of the diagonal strip and changing the timestep. Also we can 
ask ourselves how carefully is the initial
state prepared. For an exact two-point function we know that $G_{aa}(0)=-0.5i$. However, the iterations of the real-time DS equations have $G_{aa}(0)=-0.5002i$. The 
error coming from this is esentially the same as in the above number.

How do we compare this result with the equilibrium phase diagram? 
In fact, we can solve Euclidean Dyson--Schwinger equation for the coupled system+bath and compare the equilibrium energy. We start from the Euclidean correlators 
in the wormhole phase, add coupling to the bath and solve the DS equations again. The value of the 
energy we obtained this way is(again in units where $J=0.5$):
\beq
E_{\rm DS}=-0.05276(3)
\label{eta002energy_ds}
\eeq
It is represented as  green dot in Figure \ref{check001}.
The uncertainty is estimated by  changing the number of discretization points and imposing different cutoffs for the iteration procedure.
We see an agreement with \nref{eta0012_energy} within one standard deviation. This suggests that we indeed reached the wormhole.

\begin{figure}[!ht]
\centering
\includegraphics[scale=0.5]{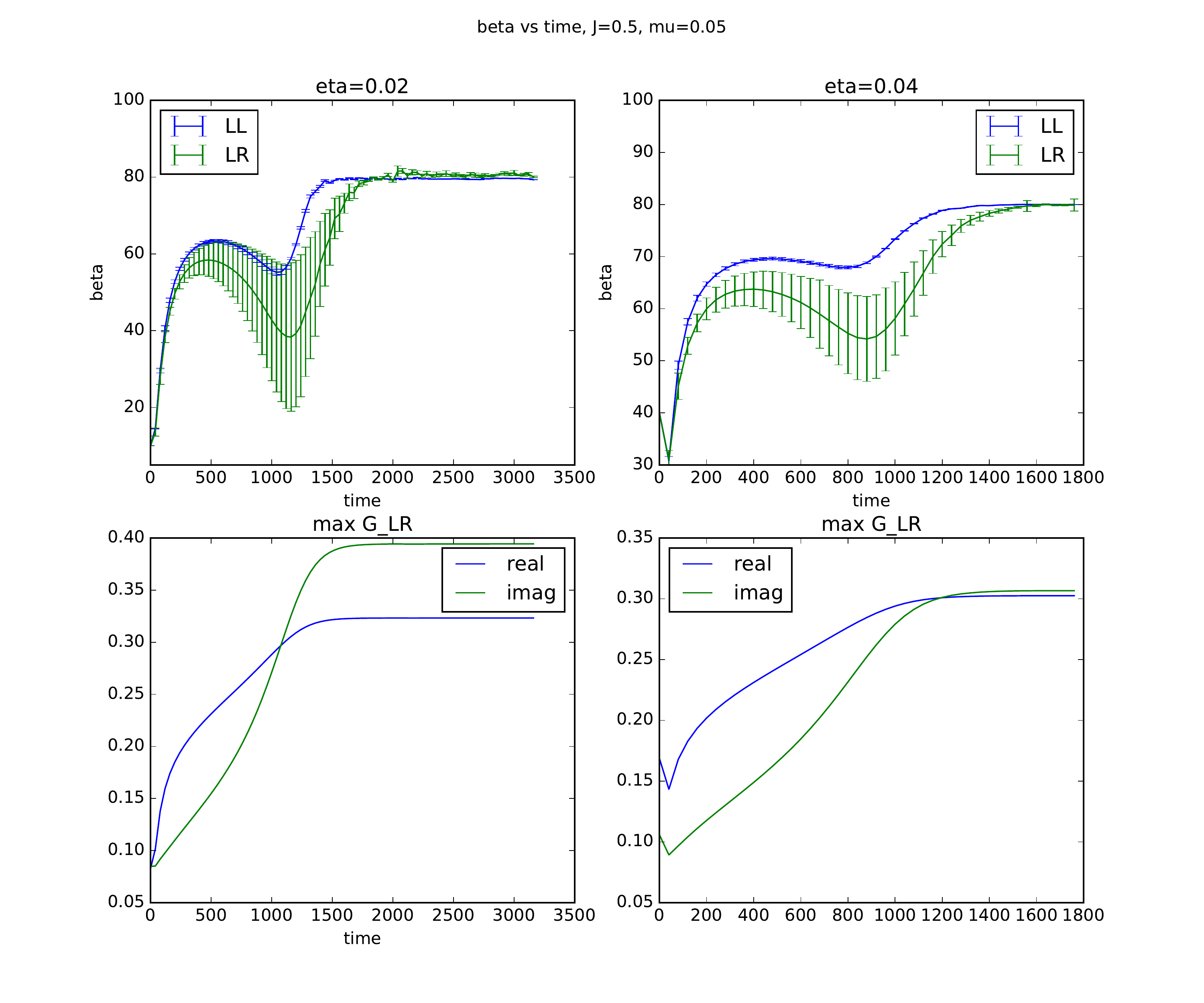}
\minipage{0.8\textwidth}
\caption{\small (Upper part) $\beta$ as a function of time for $\eta=0.02$ and $0.04$. The error bars are taken from the covariance matrix, so the meaningful quantity is the relative
size between them.
(Lower part) the maximum of value of $G_{LR}$ for the same values of $\eta$. }
\label{fig:betas}
\endminipage
\end{figure}

To understand what happens near the transition we can look at the maximum value of $G_{LR}$, see Figure \ref{fig:betas} (lower part). We can notice that,  
during the transition through the unstable phase,  the imaginary value of the correlator (which is proportional to the anticommutator) rapidly grows, 
indicating the growth in the information transmission rate.

Another thing we can see is that coupling to a bath generically thwarts the information transmission between the sides; for larger bath coupling $\eta$ the ratio
$\Im G_{LR}/\Re G_{LR}$ is smaller.

Figure \ref{check001} shows that the system is more or less following the thermodynamic curve. We see that the temperature and energy are smooth everywhere and the
transition goes through a phase with negative heat capacity, where the energy decreases and the temperature increases. 

To check whether the system remains thermal at all times we performed an additional check. Using the fluctuation dissipation theorem (FDT) 
we can find the temperature two ways: from $LL$ correlator and the  $LR$ correlator, using \nref{tanh}.
The result is shown on Figure \ref{fig:betas} (upper part). We see that $LL$ correlator is very close to thermal and the curve shows a clear period of temperature increase. In contrast,   the temperature extracted from the  $LR$ correlator has big errorbars. This means that the $LR$ correlator has larger deviations from precise thermality, and larger violations to the relation \nref{tanh}. 
We think that this happens for the following reason. 
We couple each SYK system to its own bath, so we have two uncorrelated baths. This introduces incoherence to the system,
which can be seen on Figure \ref{fig:betas}: $G_{LR}$ is decreases when we increase $\eta$. Moreover we are in the regime of small $\mu$,  which means that the two sides
interact weakly,  so the equilibration time for $G_{LR}$ is much bigger than for $G_{LL}$. Therefore we expect that the error bars for $LR$ temperature are big because
our system-bath coupling is too big. And, if we managed to cool the system down more slowly, then the $LR$ correlators would remain thermal. Unfortunately, with our limited computer resources   we could not go below $\eta=0.01$. So we do not have a clear evidence for this interpretation.

\subsection{Time to form the wormhole }
\label{sec:est_time}

In this subsection we provide analytic estimates for the time it takes to form the wormhole. 
We will first estimate the time it takes to reach $\T2BH$ and then the time it takes to 
reach from there to $\TWH$. 
At this point we basically have a cold wormhole, so we will consider it to be already formed. 
We could also consider it formed once we reach $\T2BH$ and we start moving on the hot wormhole region.

In order to estimate these times we need an expression for the rate of energy 
emission into the bath. 
For a general coupling between a system and a large bath the energy loss rate can be written as 
 (see \cite{syk_bath} for more discussion)
\be
\label{eq:flux}
{ d E \over dt } =2 i \eta \int_{-\infty}^{+\infty} d \tt \ 
\partial_{\tt} G_{LL} \l { t + \tt \over 2} , { t - \tt \over 2 } \r G_B(\tt )^3
\ee
where factor of 2 comes from having both sides of coupled to a bath. In the above integral,  we can replace the system two point function by the thermal one at the instantaneous temperature, assuming that the temperature varies slowly. 
Moreover, for SYK at low temperatures
we can simply use conformal approximations for two-point functions.

If the bath temperature
is much lower, than the system's temperature we can approximate bath Green's function by the zero temperature one
\be
G_{B} = b \frac{1}{\sqrt{J_B i \l t- i \ep  \r}}
\ee

\subsubsection{Reaching $\T2BH$}
In the two black hole holes phase we assume that $G_{LL}$ is approximated by a single SYK thermal two-point function
\beq
 G_{LL}(t)= b\frac{\sqrt{\pi}}{\sqrt{J \beta i \sinh \frac{\pi(t-i \ep)}{\beta}}}
\eeq
Plugging these into 
(\ref{eq:flux}) we see
that the answer is determined by dimensional analysis up to an irrelevant numerical coefficient\footnote{We refer to
\cite{syk_bath} for  the numerical coefficient.}:
\beq \la{EnFlux}
\frac{dE}{dt} =  - \const  \frac{\eta}{J_B^{3/2} \sqrt{J}\beta^2 }
\eeq
Differentiating the energy expression  (\ref{EnSYK}) with respect to the time,  and solving a simple differential
equation for $\beta(t)$, we find that it grows exponentially
\beq
\beta(t) \sim \beta_{\rm init} \exp \l \const \frac{\eta \sqrt{J} t}{J_B^{3/2}} \r
\eeq
So that the time $\tBH$ needed to go through the two black hole phase and reach $\T2BH \sim \mu^2/J$ 
depends only logarithmically on $\mu$:
\beq
\label{t_bh}
\tBH \sim \frac{J_B^{3/2}}{\sqrt{J} \eta} \log \frac{J}{\mu^2 \beta_{\rm init}}
\eeq

\subsubsection{Reaching $\TWH$}

Now let us calculate time $\tWH$ which is needed to go through the hot wormhole phase and reach $\TWH$.
To this end we will employ some results from the end of Section \ref{sec:sch_low} about Schwarzian. 
This Schwarzian approximation
breaks for very hot wormholes with temperature of order $\T2BH$, but holds for lower temperatures. 

Suppressing the numerical
coefficients, $G_{LL}$ in this regime is given by:
\beq
\label{G_LL_hot}
G_{LL}(t) \propto  \frac{\sqrt{\tpr}}{\sqrt{J \sin \l \tpr (t-i \ep) \r}}
\eeq
where $\tpr(t)$ is determined by the solution of eq. (\ref{saddle}). The hot wormhole(unstable branch) is characterized
by having $\tpr\ll 1$, so eq. (\ref{saddle}) can be 
simplified by neglecting Schwarzian kinetic term (the first term in eq. (\ref{saddle}))
\beq
\label{approx_saddle}
\exp \l -\tpr \beta/4 \r \sim \frac{\mu}{\tpr^{1/2}}
\eeq
This approximation breaks down near $T \sim \TWH$, so we further assume that we use this approximation for temperatures
which are slightly below or of order $\TWH$.

From this equation, up to a logarithmic term in $\mu$, $\tpr$ and $\beta$ are related by 
$ \tpr \sim  1/\beta $. 
Using  (\ref{approx_saddle}), the  energy (\ref{saddle_energy}) can be written as 
\beq
\label{E_hot}
E - 2E_{\rm 0,SYK} \propto  - \frac{\mu}{\sqrt{J \beta}}
\eeq
We see that the heat capacity is negative 
\beq
C_{\rm hot\ wormhole} = \frac{dE}{dT} = -\beta^2 \frac{dE}{d \beta} \propto  \frac{\mu \sqrt{\beta}}{\sqrt{J}} 
\eeq
We can compare the absolute value of this expression with the heat capacity of a regular SYK
$C_{\rm SYK} \sim \frac{1}{J \beta}$  
\beq
\frac{|C_{\rm hot\ wormhole}|}{C_{\rm SYK}} \propto \mu \sqrt{J} \beta^{3/2} \propto \left( { \TWH \over T } \right)^{3/2} 
\eeq
This ratio is much bigger than 1 for $\beta \gg 1/(J^{1/3} \mu^{2/3}) \sim \beta_{\rm WH}$. So apart from the
region close to $\TWH$ the hot wormhole has a large  negative heat capacity, compared to a single SYK model at the same temperature.

The energy flux can be computed using the expression (\ref{G_LL_hot}) for $G_{LL}$. The result is
again determined by scale symmetry and it is again proportional to $\tpr^2 \sim 1/\beta^2$ as in \nref{EnFlux}. 
However,  because of the big negative heat capacity, the time it takes to go through this region is much longer than  \nref{t_bh}.
Solving for $\beta(t)$ we get 
\beq
\beta_0^{3/2} - \beta_1^{3/2} = \frac{\eta}{\mu J_B^{3/2}}\l t_1 - t_0 \r
\eeq
In our case we start from $\beta_0 \sim \bet2BH$ and end with $\beta_1 \sim \betWH$.
Since $\bet2BH \gg \betWH$ the overall time length is mostly determined by the region near $\T2BH$:
\beq
\label{t_wh}
\tWH \sim \frac{J^{3/2} J_B^{3/2}}{\eta \mu^2}
\eeq

This timescale is much larger than $\tWH$, \nref{t_bh},
 which scaled only logarithmically in $\mu$. Moreover, it is mostly determined
by the region near $\T2BH$, which is where the approximation is breaking down. So \nref{t_wh} should only be viewed as an order of magnitude estimate. 

Our  numerical results for $\beta(t)$ on Figures \ref{comparison_full} and \ref{fig:betas}
seem to qualitatively support these conclusions. Notice that, as expected,  the times are inversely proportional to the coupling 
to the bath $\eta$.

\section{Two coupled black holes in gravity }
\label{sec:gravity}

The low energy description of the SYK model has some features in common with certain two dimensional theories of gravity. 
In this section, we   study a similar problem in a gravitational theory in order to compare to the answers we found above. 

We consider a Jackiw-Teitelboim theory of gravity coupled to matter, see 
\cite{cft_breaking,Jensen:2016pah,Engelsoy:2016xyb} for details. 
This gravity theory describes a two dimensional black hole with an $AdS_2$ geometry. The $AdS_2$ space has a boundary. 
We  consider a system containing two such black hole exteriors and we  introduce a coupling for the two dimensional matter fields propagating in the bulk. We assume that have $N$ such matter fields. 
Let us say that $\chi$ is a matter field with a certain mass $m$ in the bulk and quantized with Neumann
 boundary conditions so that its dimension is $\Delta$, with $\Delta < 1/2$. 
 We couple their boundary values through a term, for each field,  
\be \la{IntGra} 
 S =  i  \tilde \mu   \int du \chi_L(u) \chi_R(u)    ~,~~~~~~~~~~    \tilde \mu =      { \mu \over J^{2\Delta } }
\ee
were we imagine that $J^{-1} $ is related to a cutoff in the radial $AdS_2$ direction\footnote{With the $AdS$ metric $ds^2 = (dx^2 + dz^2)/z^2$, this is the cutoff at $z= \epsilon$, and we are defining $J = 1/\epsilon$.   } and $u$ is the physical boundary time.

\subsection{High temperature phase} 

\begin{figure}[ht!]
\centering
\includegraphics[scale=0.8]{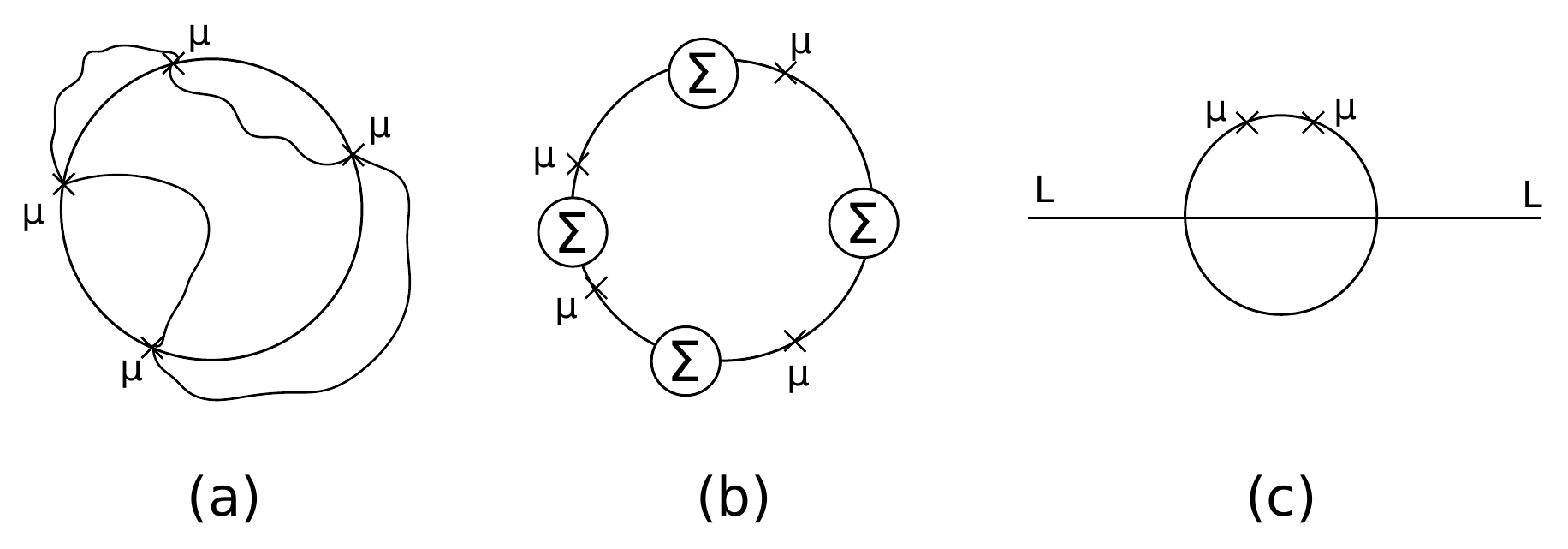} 
\minipage{0.8\textwidth}
\centering
\caption{\small (a) Field theory diagrams when we add an interaction term $i\tilde{\mu} \chi_L \chi_R$ at the
boundary of two disks. Here we picture  one disk and the second as the outside. So $\chi^L \chi^L$ propagators are the ones inside the disk and $\chi^R \chi^R$ are outside the disk.   (b) Diagrams in the two coupled  SYK model 
that reproduce the field theory diagrams in (a). (c) An example of a diagram in the two coupled 
SYK model  that is not present in field theory.}
\label{fig:diagrams}
\endminipage
\end{figure}

We now consider the high temperature phase where in Euclidean space we have two separate disks that are connected through 
the  interaction \nref{IntGra}. Concentrating on the matter system, this interaction is easy to analyze because the full matter theory is just quadratic. In principle, we also need to consider the effects of gravity, and we will discuss them later. 
This interaction, \nref{IntGra},
 leads to the Feynman diagrams in Figure \ref{fig:diagrams}(a), which can
be easily summed, as we explain below. 
 Since the interaction is relevant,   it  becomes important at low temperatures. For sufficiently low temperatures,  the net effect is to change the boundary conditions for the bulk fermions $\chi$ from Neumann  to Dirichlet. Namely, at low temperatures we get   two decoupled disks with Dirichlet boundary conditions for bulk fermions.   
We now discuss this more explicitly. 

When $\tilde \mu=0$ we have two separate disks and the matter partition function is just given by $Z_N^2$, namely the square of the partition function of a fermion with Neumann boundary conditions. 
Starting from this state we can now sum the diagrams in Figure \nref{fig:diagrams}(a). For each fermion field, we get 
\be
Z_{\mu} = Z_N^2 \exp\left\{ \half Tr[ \log (1 + i \tilde \mu G) (1-i \tilde \mu G) ] \right\} = { Z_N^2 \  \left[ \det (1 + \tilde \mu^2 G_\Delta^2 ) \right]^{\half }  }
\ee
where we think of $G(u_1,u_2) \propto [\sin { u_{1} - u_2 \over 2} ]^{-2\Delta } $ as a matrix with indices $u_1$, $u_2$. 
We have set $\beta = 2\pi$ for simplicity and we will restore it later. 

For large $\tilde{\mu}$ we find that the partition function gets an additional factor of the determinant of $G_\Delta$. 
It turns out that this produces the Dirichlet partition function \cite{KitaevRecent,Maldacena:2019cbz}  
\be \la{NDdet}
Z_{\tilde \mu \gg 1} = Z_N^2   \det G_\Delta    e^{ - \epsilon(\tilde \mu ) \beta } \propto Z_D^2 ~,~~~~~~~~ \epsilon(\tilde \mu ) \propto - ({\tilde \mu   })^{ 1 \over 1 - 2 \Delta } 
\ee
 In the last equality we neglected the energy contribution, since we will be focusing on the ground state entropy contributions. 
 
The conclusion is that if we start out with two disks with Neumann boundary conditions, after we turn on the relevant perturbation \nref{IntGra}, for very low temperatures we get two decoupled disks again but with Dirichlet boundary conditions. This implies that for very low temperatures,    the bulk fermion would be dual to an operator of dimension $1-\Delta$. 

Restoring the factors of $\beta$,  this transition happens  at $\bet2BH$ given by 
\be \la{TransiT}
1 \sim  \tilde \mu \bet2BH^{ 1 - 2\Delta } 
~,~~~~~~~~~~~{\rm or } ~~~~~~~ \T2BH \propto \left( { \mu \over J^{ 2 \Delta } } \right)^{ 1 \over 1 - 2 \Delta }  
\ee
For $q=4$ this reproduces \nref{T1_est}.  
This is not surprising because we were summing the same type of diagrams.
However, in the gravity case these are all the diagrams, so we can study the whole flow. The new IR fixed point simply corresponds to flipping the boundary conditions to Dirichlet. So nothing too dramatic happens in the gravity solution when we go to 
temperatures lower than the  temperature $\T2BH$ in \nref{TransiT}, except that the change in the boundary conditions will change the value of the ground state entropy. 

In this discussion,  we have ignored the dynamics of gravity. In principle, we could wonder 
whether we should consider non-trivial solutions of the Schwarzian theory. 
If we assume that the solution is invariant under translations for each disk, then, 
up to gauge symmetries of the Schwarzian theory,  the only solution is the usual one. In Appendix \ref{sch:1loop}, we examine whether  non-constant Schwarzian modes could lower the action. We find that they do not, at least in the approximation we considered. 
In our analysis we assumed that gravity is classical, 
 which is correct if 
$\phi_r/\beta \gg 1$. Here $\phi_r$ is the JT gravity analog of the coefficient of the Schwarzian, the analog of 
$N \alpha_S/{\cal J}$. We have also assumed that we have a relatively low number of bulk quantum fields so that the effects of integrating them out does not significantly change the value of $S_0$, the ground state entropy. This is   the regime where the 
gravity theory is simplest. As we will discuss below,  the SYK model is different in this respect.

\subsection{Low temperature phase} 

At very low temperatures  the coupling \nref{IntGra}  leads to the formation of a wormhole
\cite{MQ}. 
This is identical to the small $\mu$ coupled SYK model analysis of 
 section \ref{sec:sch_low}, since the effects of gravity can also be described in terms of the 
Schwarzian mode.

When we decrease the temperature along the negative specific heat region (the hot wormhole phase),    the wormhole is getting longer and longer. Or, more precisely, there is a larger redshift factor between the boundary and the center of the wormhole. 
Then,  
the interaction, which is a relevant deformation, 
becomes stronger. When we considered the problem for the disks, we found that for strong interactions   
we get an effective change in boundary conditions from Neumann to Dirichlet. 
Here we expect the same phenomenon   when $\tpr$ is becomes  
\be
 {\tpr}  \propto  \T2BH  \propto \left( { \mu \over J^{2\Delta } } \right)^{ 1 \over 1 - 2 \Delta } 
\ee
where $\tpr$ is the variable in \nref{saddle}, which is proportional to the value of the redshift factor at the center of the wormhole. 
In other words, $\tpr$ becomes of the order of the temperature $\T2BH$ 
in \nref{TransiT}. We refer to Appendix \ref{app:flow} for details.
At this value of $\tpr$ the wormhole is so long that the 
approximations used in deriving \nref{Constr} are no longer valid. 
Interestingly, due to \nref{UnstT}, 
this happens also at a temperature of the order of $\T2BH$,  which is the temperature where the two disk solutions starts being corrected.  
This might appear as a coincidence, but it is not. In the hot wormhole phase we find that the temperature sets the value of 
$\tpr$ and thus the amount of RG flow that the relevant left-right interaction undergoes. Therefore, this interaction becomes relevant at the same place. 

This statement can be further verified by checking whether the hot wormhole thermodynamic curve (red curve in figure \ref{sch_comp})
will join with the two black hole phase (blue curve in figure \ref{sch_comp}) at $T=\T2BH$. In the hot wormhole phase
the energy is given by  (\ref{E_hot}) and in the two black hole phase by 
(\ref{delta_E}). 
Indeed, the two curves join at $T \sim \T2BH$.

In a gravity theory with a relatively low number of fields, we expect that after $\T2BH $  the wormhole phase might not exist any longer.

One conclusion is that, in a general JT gravity theory plus matter, 
 we do {\it not} seem to be able to easily join the   high temperature phase and the hot wormhole phase. 
 This is mysterious in the gravity theory because it involves a topology change. 
  Of course, the low temperature phase and the hot wormhole  phase are connected smoothly at $\TWH$ in a region where we can trust the   wormhole analysis in the Schwarzian approximation, as discussed near \nref{Constr}.

\subsection{Comparison with the SYK model } 

In the SYK model, the addition of the interaction corrects the original diagrams by inserting $\mu$ terms in the propagators. If we insert them outside the self energy correction (the $\Sigma$ bubble), then we get diagrams which are identical to the ones we discussed in gravity, see Figure \ref{fig:diagrams}(b). However, in the SYK model we can also 
insert $\mu$ inside the self energy corrections, these are new diagrams that are not present in the gravity discussion, 
see Figure \nref{fig:diagrams}(c).

 As we mentioned above the region of the phase diagram near temperatures $T\sim \T2BH$ is different in a generic JT gravity theory plus matter than in SYK.  However, we can consider the folloing gravitations model that would look more {\it qualitatively} similar to the SYK model. 
  
First we note that the SYK ground state entropy is given by $S_0 = N s_0$, where $s_0$ is given by \cite{kitaev,KitaevRecent}
\be \la{EntSYK}
 e^{s_0} =( \det G_{\Delta} )^{-\half } = { Z_N \over Z_D }   ~,~~~~~~~~~ {\rm for ~one~ Majorana~fermion}
 \ee
 where the first equality follows from the usual $G,\Sigma$ action at low energies. The second equality was  mentioned   in \nref{NDdet}. 
 This implies that if we want to describe the SYK model in terms of JT gravity, we should think that when the fermions have Dirichlet boundary conditions, the net entropy, or value of the topological terms in the action should be zero, $\phi_0=0$. Then the actual value of the ground state entropy of the usual, single boundary SYK  model, \nref{EntSYK}, is simply given by the contribution of changing the boundary condition for the bulk fields from Dirichlet to Neumann \cite{KitaevRecent}. 
 
 Returning now to the coupled model and starting from the high temperature phase, we see that when we reach the 
temperature $\T2BH$ we are changing to a Dirichlet boundary condition. This means that the total $S_0$ now becomes zero, which implies that the topology change is easy. 
 Similarly, if we start from the canonically unstable wormhole phase and approach $\T2BH$, we also see a change in the boundary conditions so that $S_0$ again becomes zero and topology change is easy. 
 So we can join the two phases with a change in topology at $\T2BH$. In this way we can { \it qualitatively} understand the transition. We have given evidence that this is a smooth transition in the coupled SYK model. What we are discussing here is just a cartoon for a gravity picture of what is happening. 
  
 We also see why SYK is different than a generic JT gravity theory with a smaller number of fields. In such gravity theories the flow from Neuman to Dirichlet would not change $S_0$ by too much and the topology change remains suppressed. 
 For this reason we have not been able to see a 
general mechanism for the transition that would also work in 
more general gravity theories, such as the Standard Model in the presence of magnetically charged black holes as discussed in 
 \cite{MMP}.

\section{Conclusion}

In this paper we studied the approach to the ground state of the two coupled SYK models \cite{MQ}. 
We first discussed the equilibrium thermodynamics picture. 
In the microcannonical ensemble we expect a continuous picture with no phase 
transition. As the energy decreases, the temperature decreases up to a value $\T2BH$ where 
the system looks like two separate thermal SYK models with a small coupling. 
At $\T2BH$ this coupling becomes strong and the system transitions to 
a ``hot wormhole" phase with negative specific heat. Now the temperature increases 
up to $\TWH$ and then the wormhole becomes cooler and shorter, and the specific heat becomes positive again. 
This whole picture can be understood using simple analytic approximations, 
except for the transition region at $T\sim \T2BH$. 

We found that the real time evolution looked as if the system is following the above equilibrium phase diagram. Unfortunately, for the parameters we could use in our numerical computation, we could not trust quantitatively the simple analytic approximations. However, these gave a qualitatively correct answer. The system remained near thermal equilibrium as it cooled down, except for some deviations in the $G_{LR}$ correlator, which we think should disappear if we were to cool more slowly. 

The conclusion is that, starting with a generic state of the 
two coupled SYK model, we can find the ground state by coupling the system to the bath and cooling it  down. In particular, the system does not get stuck in a metastable state. This provides a feasible way to produce a state close to the TFD. 

We also computed the time to form the wormhole. Most of the time is spent near the region with $T \sim \T2BH $. 

One of our goals was to extract some general lessons for wormhole formation in gravity. Unfortunately, the SYK model seems to be special, and its special features becomes manifest in the ease by which we can connect the two black hole phase with the hot wormhole phase near $T \sim \T2BH$. 
These two phases do not seem to be so easy to connect in more general theories of gravity. We  qualitatively explained  why topology is  simpler in a gravity theory that is similar to the SYK model, but harder in a more general theory of gravity. 

Nevertheless we cannot say how hard forming a wormhole would be in a more general theory of gravity, such as the one describing the wormholes in \cite{MMP}. It seems hard, but maybe there is an ``easy'' pathway to form it. It would be interesting to answer this question. 

We expect that this article would be relevant for efforts that try to do it using the SYK model, see the proposal in  \cite{Chen_2018}, for example.

\section*{Acknowledgments}
We are indebted to Ahmed Almheiri and Brian Swingle for describing to us how to couple SYK to a bath and numerous discussion about the thermalization in SYK.
Also we are grateful to Ksenia~Bulycheva, Yiming~Chen, Alexander~Gorsky, Luca~Iliesiu, Alex~Kamenev, Alexei~Kitaev, Igor~Klebanov, 
Fedor~Popov, Xiaoliang Qi, Nick~Sukhov, Grisha~Tarnopolsky and Zhenbin~Yang for discussions.
J.M. is supported in part by U.S. Department of Energy grant DE-SC0009988 and by the Simons Foundation grant 385600.

\appendix

\section{Perturbation theory in $\mu$}
\label{app:delta_glr}
Let us start from computing $\mu$ correction to $G_{LR}$.
We need to compute
\begin{align}
\Delta G_{LR}(\tau_1-\tau_2) = i \mu \int_0^\beta d\tau \ G_{LL}(\tau_1-\tau) G_{RR}(\tau-\tau_2)  = \nonumber \\
= i \frac{\mu b^2 \pi}{J \beta} \int_0^\beta d\tau \ \sgn(\tau_1 - \tau) \sgn(\tau-\tau_2) \frac{1}{\sqrt{\sin \l \frac{\pi |\tau_1-\tau|}{\beta} \r}} \frac{1}{\sqrt{\sin \l \frac{\pi |\tau-\tau_2|}{\beta} \r}} 
\end{align}
Because of the translational invariance along the thermal circle the answer depends on $\tau_1-\tau_2$ only. So one can put $\tau_2=0$.
After changing the variables to $x=\tan \l \pi \tau/\beta \r $ we have the following integral:
\beq
i (1+x_1^2)^{1/4} \times \frac{\mu b^2}{J} \int_{-\infty}^{\infty} dx \ \sgn(x_1-x) \sgn(x) \frac{1}{\sqrt{|x|} \sqrt{|x_1-x|} \sqrt{1+x^2}}
\eeq

This integral can be computed analytically. Because of the $\sgn$ functions there are three integration domains. So the answer has three parts:
\beq
\Delta G_{LR}(x(\tau)) = i \frac{\mu b^2}{J} (I_1-I_2+I_3) (1+x^2)^{1/4}
\label{glr_conf}
\eeq
where 
\bea
I_1 &=&\frac{2}{\sqrt{1- i x}} \l \text{EllipticK} \l 1- \frac{2x}{i+x}  \r - i \text{EllipticF}\l \frac{\pi}{4}, \frac{2x}{i+x} \r \r \nonumber \\
I_2&=&\frac{2}{(1+x^2)^{1/4}} \text{EllipticK}\l\frac{1}{2} - \frac{1}{2\sqrt{1+x^2}} \r \\ \nonumber
I_3&=& \frac{2}{x} \ {}_3 F_2 \l\ \oh,\oh,1 ; \frac{3}{4},\frac{5}{4};-\frac{1}{x^2} \r
\label{}
\eea

Now, let us compute the correction to energy.
From the path integral the leading correction to the free energy is
\beq
\label{delta_F}
-\beta \Delta F  = -\frac{\mu^2 \beta}{2} \int_0^\beta d\tau \ G_{LL}(\tau) G_{RR}(\tau)
\eeq

We can recover the integral by taking  $\tau \ra 0$ limit in the conformal answer (\ref{glr_conf}) for $G_{LR}$.
Unfortunatelt it produces a logarithmic UV divergence which we cut at $\tau=1/J$:
\beq
\Delta F = \frac{1}{2 \sqrt{4 \pi}} \frac{\mu^2}{J} \l 2 \log \frac{\pi}{J \beta} - c_1 \r
\eeq
where $c_1$ is the cut-off dependent constant. We can not find it from the conformal perturbation theory, because it is an effective low-energy theory with a build-in UV cutoff of
order $J$. From the above expression we read off the energy correction:
\beq
\Delta E = \frac{1}{2 \sqrt{4 \pi}} \frac{\mu^2}{J} \l 2 \log \frac{\pi}{J \beta} -2 - c_1  \r
\eeq
Also notice that the constant $-c_1-2$ is not simply a correction to the ground state energy. 
As we mentioned in the main text, perturbation theory in $\mu$ breaks down
at large $\bet2BH \sim J/\mu^2$, so we can not take the limit $\beta \ra \infty$ 
in this expression. This is signalled by the presence of the logarithm. 
This divergence has IR nature, and it is not caused by using the conformal answer in the integral (\ref{delta_F}).

To extract $c_1$ we can compute integral (\ref{delta_F}) using the numerically obtained $G_{LL}$. 
We find that $c_1 = 1.66(1)$. This agrees very well with the actual numerical result for the energy - 
Figure \ref{sch_comp}.

\section{Checking whether Schwarzian fluctuations are stable }
\label{sch:1loop}
Let us start from doing a 1-loop calculation for Schwarzian first.
We again assume that we in the phase with two separate black holes. Then the action will involve two Schwarzian terms plus the interaction piece.
For a moment we can imagine that instead of a simple interaction $\mu \psi_L \psi_R$ we have a term involving composite operators $ i^{F} \mu \Oc_L \Oc_R$ with zero
expectation values: $\bra \Oc_{L(R)} \ket=0$. $F$ is the $\Oc_{L(R)}$ fermionic number. The Schwarzians interact because 
Schwarzians gravitationally dress $\Oc_{L(R)}$ by time reparametrizations.

In the perturbation theory the leading contribution is of order $\mu^2$:
\begin{align}
\int D \psi_L D \psi_R \ \exp \l -S_L-S_R - i^F \mu \int d u \ \Oc_L \Oc_R \r = \\ \nonumber 
= \int D \psi_L D \psi_R  \ \exp \l -S_L - S_R - 
\mu^2 \int du_1 du_2 \bra \Oc_L \Oc_L \ket \bra  \Oc_R \Oc_R \ket \r
\end{align}
Explicitly the action is:
\begin{align}
\label{sch:action}
S = -\int du \ \frac{N \al_S}{\Jc} \{f_L,u\} - \int du \ \frac{N \al_S}{\Jc} \{ f_R,u \} -  \\ \nonumber
N \frac{\mu^2 c^2_{\Oc}}{J^{4 \Delta}} \int du_1 du_2 \l \frac{f'_L(u_1) f'_L(u_2)}{(f_L(u_1)-f_L(u_2))^2} \r^\Delta \l \frac{f'_R(u_1) f'_R(u_2)}{(f_R(u_1)-f_R(u_2))^2} \r^\Delta
\end{align}
where $\Delta$ is the dimension of $\Oc_{L(R)}$, for $\psi_L$ it is $\Delta=\frac{1}{4}$ and $\{f,u\}$ denotes Schwarzian derivative:
\beq
\{f(u),u \} = \frac{f'''}{f'}-\frac{3}{2} \l \frac{f''}{f'} \r^2
\eeq

Finite temperature solution without interaction reads as:
\beq
f_L = f_R = \tan \frac{\pi u}{\beta}
\eeq
We can perturb it by $\ep_L,\ep_R$:
\begin{align}
\label{eq:thermal}
f_L = \tan \l  \frac{\pi u}{\beta} + \ep^L(u) \r \\
f_R = \tan \l  \frac{\pi u}{\beta} + \ep^R(u) \r \\
\label{}
\end{align}
and to see whether the two black hole system has a perturbative instability at some temperature.
For simplicity we can put $\beta=\pi$ and expand $\ep$ in Fourier modes:
\beq
\label{eq:fourier}
\ep = \sum_{n=-\infty}^{+\infty} \ep_n e^{2 i n u}
\eeq

Before doing an actual computation, let us stop and explain 
why Schwarzian does not contribute classicaly here and at higher loops. By classically we mean that its contribution
is suppressed by $1/N$. Since we are in a thermal
state (\ref{eq:thermal}) is the a translation symmetry along the Euclidean time $u$. This is why after expanding in
Fourier modes (\ref{eq:fourier}) we will not have terms linear in $\ep_n$. It means that the thermal solution
(\ref{eq:thermal}) is still a classical solution of Schwarzian equations of motion even with complicated non-local
interaction induced by loops. Since we have an overall $N$ in front of the action, integrating out $\ep_n$ will
lead to a subleading correction.

Let us return to the actual 1-loop calculation. The only subtlety is that one has to be careful with the time ordering, since 
the denominator involves 
\beq
\l \sin^2 \l u_1-u_2 + \ep^L(u_1) - \ep^L(u_2) \r \r^{\Delta} = 
\sin^{2 \Delta} | u_1-u_2 + \ep^L(u_1) - \ep^L(u_2)|
\eeq
The result for the
marginal deformation $\Delta=1/2$:
\beq
S/N=\frac{\al_S}{\Jc \beta} \sum^{+\infty}_{n=2} \ep^{l,r}_{-n} \l n^4-n^2 \r \ep^{L,R}_n+ \frac{c_{\Delta}^2 \mu^2 \beta^2}{(J\beta)^2}
\l 8 \pi^2 |\ep^L_2-\ep^R_2|^2 + 32 \pi^2 |\ep^L_3-\ep^R_3|^2 + 80 \pi^2 |\ep^L_4-\ep^R_4|^2 \r + \dots
\eeq
and the coefficients tend to grow. One can also evaluate non-quadratic terms. Below are the first three. All of them have positive coefficients too:
\begin{align}
+28\pi^2 |\ep^L_2-\ep^R_2|^4 + 224 \pi^2 |\ep^L_3-\ep^R_3|^4 + 952 \pi^2 |\ep^L_3-\ep^R_3|^4 + \dots \\ \nonumber
+ \frac{2860 \pi^2}{9} |\ep^L_2-\ep^R_2|^6 + \dots
\end{align}

For the case of relevant deformation $\mu \psi_L \psi_R$ with $\Delta=1/4$ the results are similar. The interaction term has the expansion:
\begin{align}
\frac{8 }{3} |\ep^L_2+ \ep^R_2|^2 + 8 |\ep^L_2-\ep^R_2|^2 + \frac{48}{5} |\ep^L_3+\ep^R_3|^2 + \frac{80 }{3} |\ep^L_3-\ep^R_3|^2 + \dots \\ \nonumber
+\frac{304}{15} |\ep^L_2+\ep^R_2|^4 + \frac{4432 }{105} |\ep^L_2-\ep^R_2|^4 + \frac{7146}{55} |\ep^L_3+\ep^R_3|^4 + \frac{137018}{495} |\ep^L_3-\ep^R_3|^4 + \dots \\
+ \frac{135424 }{693} |\ep^L_2+\ep^R_2|^6 + \frac{1053952 }{2835} |\ep^L_2-\ep^R_2|^6 + \dots
\end{align}
And the coefficient in front is $b^2  \frac{\mu^2 \beta^2}{J \beta}$.

In principle, we can go to higher orders in $\mu$. Curiously, $\mu^4$ correction is negative for $\ep_2$.

\section{Lorentz time Dyson--Schwinger equation}
\label{ds:lorentz}
\subsection{Single SYK}
Let us first consider a single SYK in Euclidean time.
Then the DS equations have the form:
\begin{align}
\label{ds:euclidean}
\Sigma(\tau)= J^2 G(\tau)^3 \nonumber \\
(-i \om - \Sigma(\om)) G(\om)= 1
\end{align}

Now we want to switch to Lorentzian time. We define the Wightman function with an extra $-i$:
\beq
-i \bra \psi_i(t) \psi_i(0) \ket = G^>(t)
\eeq

As is well-known, upon the analytical continuation in time domain, the time-ordered Euclidean two-point 
function becomes the Wightman function, therefore\footnote{The minus sign is subtle: one can recover it either
from the effective action (\ref{eq:eff_action}) on the Keldysh contour  or doing a careful analytic continuation through the frequency space as was done in \cite{MS}}:
\beq
\Sigma^>(t) = - J^2 \l G^>(t) \r^3
\eeq
The other DS equation is written in the frequency space, this is why after the analytic continuation it will involve the retarded components:
\beq
G^R(\om)(\omega - \Sigma^R(\om))=1
\label{gr_sr}
\eeq
So far we have not used any information about the state we are considering. This information is needed to connect $G^>$ and $G^R$. In thermal state we can use
Fluctuation-Dissipation Theorem(FDT):
\beq \la{GpGR}
G^>(\om) =  \frac{2 i \Im G^R(\om)}{e^{-\beta \om}+1} 
\eeq

\begin{figure}[ht!]
\centering
\includegraphics[scale=0.5]{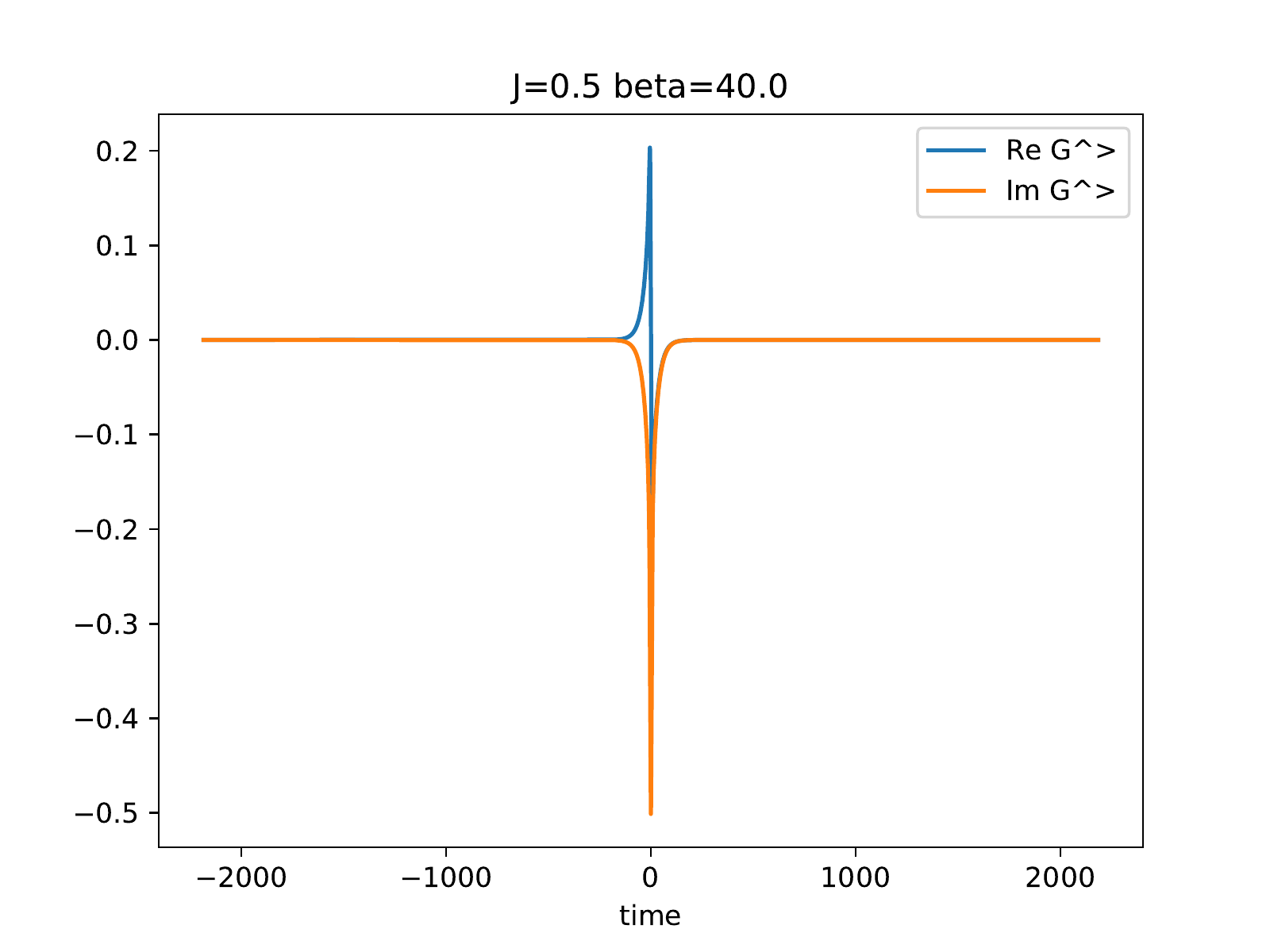}
\caption{Wightman's function $G^>$ for single $q=4$ SYK with $J=0.5,\ \beta=40$}
\label{fig:single_G}
\end{figure}

An example of how the Wightman's function look is presented on Figure \ref{fig:single_G}.

\subsection{Two coupled SYK}
Now we have 4 Green's functions. The self-energies have similar expressions:
\beq
\Sigma^>_{ab}(t) = - J^2 (G^>_{ab}(t))^3 + i \mu_{ab}
\eeq
and the equation (\ref{gr_sr}) should be understood as a matrix equation.
For the diagonal Green's functions the FDT has the same form:
\beq
G^>(\om)_{aa} =  \frac{2 i \Im G_{aa}^R(\om)}{e^{-\beta \om}+1} 
\eeq
However the imaginary part of the off-diagonal components is skew-symmetric in time, so we have:
\beq
G^>(\om)_{LR} =  \frac{\Re G_{LR}^R(\om)}{e^{-\beta \om}+1} 
\eeq

This system of equations can be solved numerically by the iteration procedure used for a single SYK model \cite{MS}.
To ensure that the iteration procedure converges to an actual solution we monitor the discrepancies of eqs. (\ref{eq:ds}):
\begin{align}
\label{eq:disc}
d_{ab}=\frac{1}{N_{\rm points}}\| \pr_\tau G_{ab}(\tau) - \Sigma_{ac} * G_{cb}(\tau) - \delta_{ab} \delta(\tau) \|^2
\end{align}
and make sure that $d_{ab} < 10^{-10}$. The typical number of discretization points is $N_{\rm points} \sim 2^{17}$.

\begin{figure}[ht!]
\centering
\includegraphics[scale=0.5]{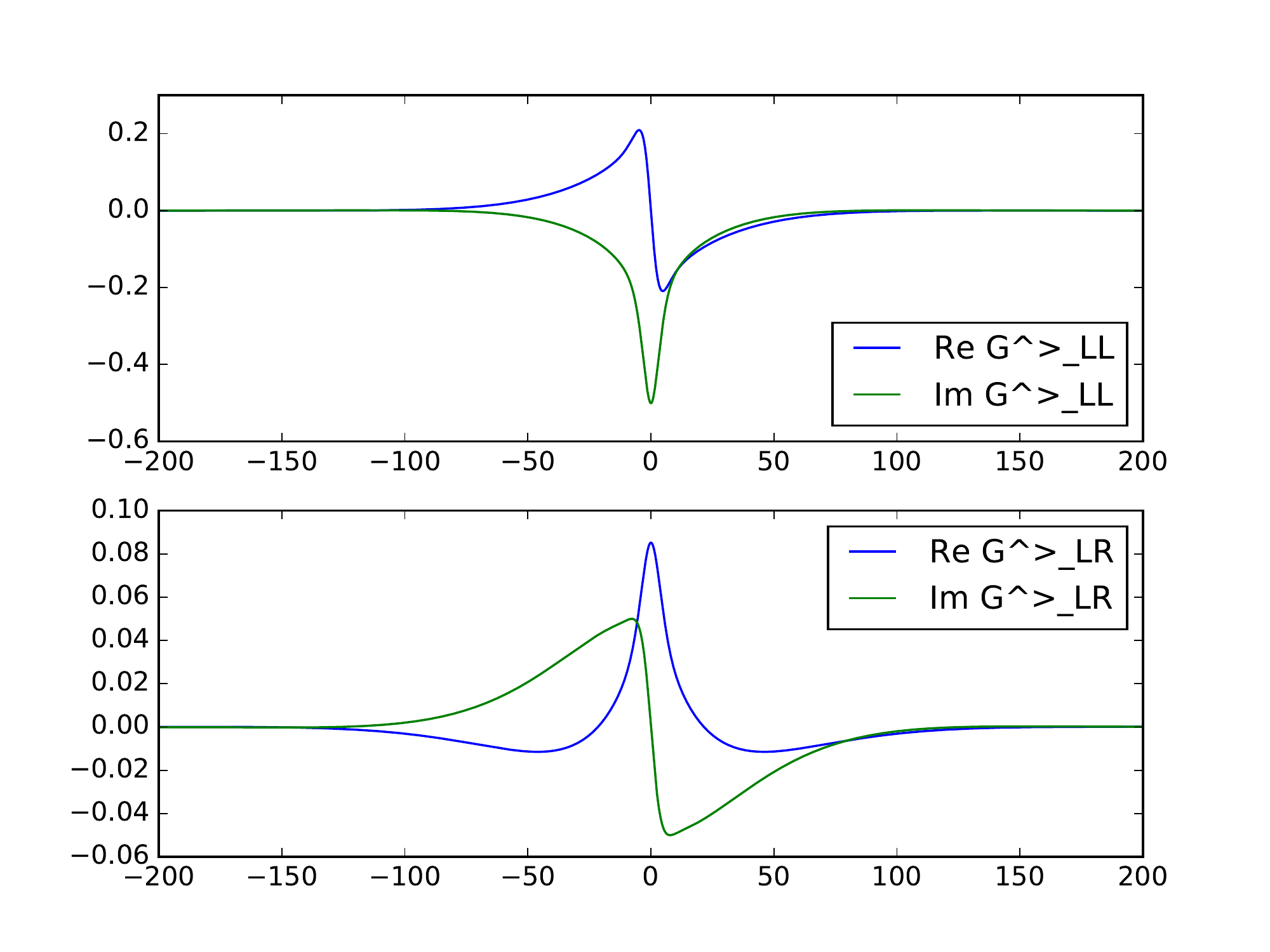}
\minipage{0.8\textwidth}
\centering
\caption{Wightman's function $G^>$ for two coupled $q=4$ SYK with $J=0.5,\ \mu=0.05,\ \beta=40$}
\label{fig:two_G}
\endminipage
\end{figure}

Before the transition the diagonal Green's function look similar to single SYK ones - Figure \ref{fig:two_G}

\section{Derivation of the effective action}
\label{eff_action_der}
Let us write down explicitly the total action for the system on the Keldysh contour $\Cc$. 
We will suppress the bath action. Bath fermions $\chi_{\alpha}$ and
$\tilde{\chi}_\alpha$, $\alpha=1,\dots,M$ belong to independend $q=4$ SYK systems with coupling $J_B$. We denote their two-point functions by $G_B$:

\beq
G_B^>(t^-_1,t^+_2) = -i \frac{1}{M} \sum_{\alpha} \bra \chi_\alpha(t^-_1) \chi_\alpha(t^+_2) \ket = -i \frac{1}{M} \sum_{\alpha} \bra \tilde{\chi}_\alpha(t^-_1) \tilde{\chi}_\alpha(t^+_2) \ket
\eeq 
The total action consists of four terms:
\beq
 S_{\rm tot} = S_{\rm kin} + S_{\rm J} + S_{\mu} + S_{\rm bath} 
\eeq
\begin{itemize}
\item $S_{\rm kin}$ is a standard kinetic term for fermions:
\beq
i S_{\rm kin} = i \int_\Cc dt \ \sum_j \l \psi^L_j \pr_t \psi^L_j + \psi^R_j \pr_t \psi^R_j  \r 
\eeq
\item $S_{\rm J}$ is SYK interaction:
\beq
i S_{\rm J} = i \int_\Cc dt \ \sum_{i<j<k<l} J_{ijkl} \l  \psi_i^L \psi_j^L \psi_k^L \psi_l^L + 
\psi_i^R \psi_j^R \psi_k^R \psi_l^R  \r
\eeq
\item $S_{\mu}$ is Maldacena--Qi interaction term:
\beq
i S_{\mu} = \int_\Cc dt \ \mu \sum_j \psi^L_j \psi^R_j  
\eeq
\item Finally $S_{\rm bath}$ is interaction with the bath:
\beq
i S_{\rm bath} = i \int_\Cc dt \  \sum_{\al_1 < \al_2 < \al_3, i}  \l V_{i}^{\al_1 \al_2 \al_3} \psi_{L,i} \chi_{\al_1} \chi_{\al_2} \chi_{\al_3} + 
\tilde{V}_{i}^{\al_1 \al_2 \al_3} \psi_{R,i} \tilde{\chi}_{\al_1} \tilde{\chi}_{\al_2} \tilde{\chi}_{\al_3}\r 
\eeq
\end{itemize}
As usual, we can integrate out the disorders leading to bi-local expression in terms of $\psi, \chi$ and $\tilde{\chi}$.
Couplings $V$ and $\tilde{V}$ are Gaussing with the variance \cite{syk_bath}:
\beq
\bra \l V_{i}^{\al_1 \al_2 \al_3} \r^2  \ket = \bra \l \tilde{V}_{i}^{\al_1 \al_2 \al_3} \r^2  \ket = \frac{3! \eta}{M^{3}} , \ \text{no sum}
\eeq 
The action can be made quadratic in fermions by introducing the largrangian multiplier $\Sigma$, which is integrated over along the imaginary axis:
\beq
i S_{\rm lag} = - \frac{N}{2} \int_\Cc dt_1 dt_2 \ \sum_{ab} \Sigma_{ab}(t_1,t_2) \l G_{ab}(t_1,t_2) - \frac{1}{N} \sum_j \psi^a_j(t_1) \psi^b_j(t_2) \r
\eeq
Note that we have an overall minus if front of the action. It is important for the equation connecting the self-energies $\Sigma$ and Green's functions.
Integrating out the fermions produces we effective action (\ref{eq:eff_action}):
\begin{align}
i \frac{S_{\rm tot}}{N}=\frac{1}{2} \Tr \log \l \omega - \Sigma_{ab}(\omega) \r - 
\sum_{ab} \int_\Cc dt_1 dt_2 \l  \frac{J^2}{8}  G_{ab}(t_1,t_2)^4  + \frac{1}{2} \Sigma_{ab}(t_1,t_2) G_{ab}(t_1,t_2) \r + \nonumber \\
+\frac{i \mu}{2} \int_\Cc dt \ \l G_{LR}(t,t) -G_{RL}(t,t) \r - \frac{\eta}{2} \int_{\Cc} dt_1 dt_2 \l G_{LL}(t_1,t_2) + G_{RR}(t_1,t_2) \r G_B^3(t_1,t_2)
\end{align}
Variation of this action with respect to $\Sigma_{ab}$ and $G_{ab}$ yield the KB equations (\ref{eq:kb}) and (\ref{def:sigma})

\section{Numerical method}
\label{numerics}

Now let us describe the numerical method for solving Kadanoff--Baym equations. Numerical solution of Kadanoff--Baym equations for SYK model was described 
previously in \cite{quench1,quench2} and our approach is essentially the same.

\begin{figure}[!h]
\centering
\includegraphics{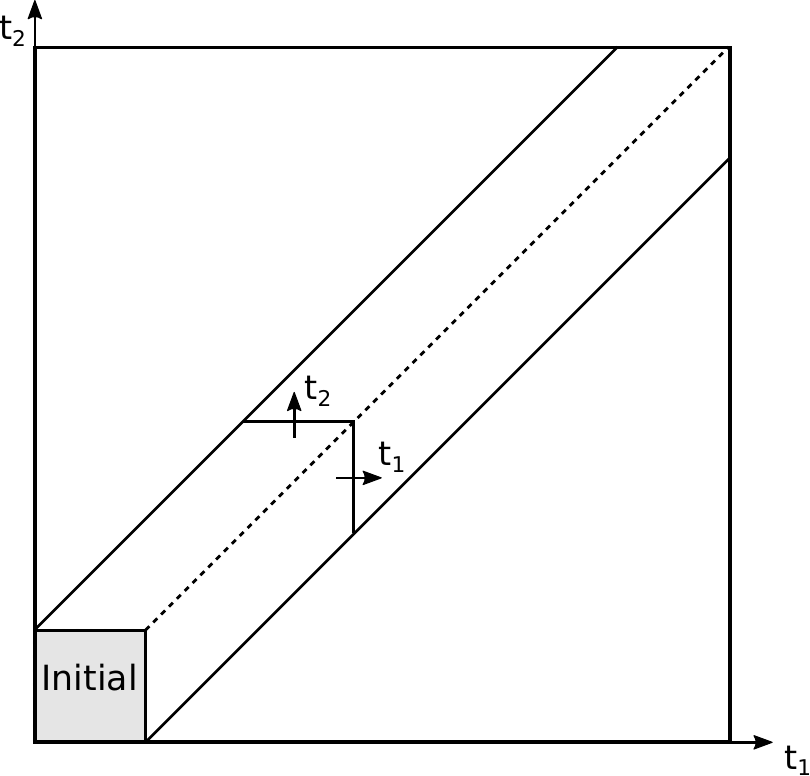}
\caption{Geometry of the $(t_1,t_2)$ plane. The initial Green functions are placed inside the gray box.}
\label{grid}
\end{figure}

We will use two-dimensional grid with uniform timestep to discretize $(t_1,t_2)$ plane. The timestep $dt$ should be much smaller than
the characteristic time-scales in SYK $1/J,1/\mu$. Since $\mu \ll J$, the $1/J$ constraint is much stricter. 
We will work with $J=0.5$ this is just a choice to fix energy units. Our time steps will be $0.2,0.1,0.05$. 
The main numerical limitation comes from the fact that the Green functions have spread $\sim \beta$, so we can 
not go to very big $\beta$, since we will have
to use a huge grid. At finite temperature the Green functions decay exponentially, so 
to greatly speed up the computation we will concentrate on the strip $|t_1-t_2| \lesssim c \beta_{\rm max}$ on the $(t_1,t_2)$ plane - Figure \ref{grid}. We will
assume that outside this strip all the Green functions are zero.
The constant $\beta_{\rm max}$ is the maximal $\beta$ in the problem at hand. In our case $\beta_{\rm max}=\beta_{B}$ - bath's beta. One can verify that one $c$ is big enough the
result of the computation does not change.

With the computation power avaliable to us, in order to keep the computation time to be of 
order of dozens of hours, $\beta$ should be less than $100$. 
This limits us to   $\mu \gtrsim 0.05$. For $\mu=0.05$ the transition beta is $\sim 61$ - see
Figure \ref{sch_comp}.

The system-bath coupling $\eta$ should be much smaller than $J^2=0.25$ so that the system remain thermal.
We will use $\eta$ in the range $0.01-0.04$. Moreover to avoid large gradients we will switch on the coupling linearly, with the switch-on time $T_{\rm switch}=20$.

Initial Green function is found by numerically solving Lorentz-time equilibrium Dyson--Schwinger equation as described in Appendix \ref{ds:lorentz}.
The bath Green function is obtained in a similar fashion. In order to avoid large speads in the bath's Green function, $\beta_B$ will be in the range $70-100$.

To compute the integral in KB equations we will use trapezoid method and for the time propagation we use predictor-corrector scheme. The same techniques have been used in
\cite{quench1,quench2}. 
For KB equations one has to be careful with propagating the Green function along the diagonal. Fortunately, for Majorana fermions there is a simple relation:
\begin{equation}
G^>_{aa}(t,t) = -\frac{i}{2}
\end{equation}
However, for the Green function obtained by numerically solving the DS equation the diagonal value is not exactly $-i/2$, so on a discrete lattice we will
just propagate this value:
\beq
G^>_{aa}(j,j)=G^>_{aa}(0,0)
\eeq
For the mixed $G_{LR}$ we do not have a simple relation like that. So we will use again the predictor-corrector scheme. The value on the diagonal can be found
by either propagating along $t_1$ or $t_2$. We will take the avarge of these results.

Let us define the ``corner slice'' Green functions as $G^>_{T,ab}(x)$:
\begin{equation} \la{Corner}
G^>_{T,ab}(x)=\theta(x) G^>_{ab}(T-x,T)+ \theta(-x) G^>_{ab}(T,T+x)
\end{equation}
Then the energy at time $t=T$ can be computed analogously to eq. (\ref{eucl_energy2}):
\beq
\frac{E}{N} = \frac{i}{4} \l \pr _x G^>_{T,LL} + \pr_x G^>_{T,RR} + \frac{i \mu}{2} G^>_{T,LR} \r,~~~~~~~~~ \text{at} \ x=0
\eeq
Computing the time derivatives using the KB equtions one arrives at the following integral form:
\beq
\frac{E}{N} =-i \frac{J^2}{4} \int_{-\infty}^{\infty} dx \sgn(x) \l G^>_{T,LL}(x)^4 + G^>_{T,RR}(x)^4 + 2 G^>_{T,LR}(x)^4 \r  + i \mu G^>_{T,LR}(0)
\label{kb_energy}
\eeq

Along the time evolution the system cools down, so formally the temperature is not well defined. However, if the cooling process is slow we might expect that 
at each point in time the short-time correlators will be thermal. In order to extract the 
temperature we need to choose the 1D slice of the Green function $G^>_{S,ab}$ to
use the Fluctuation-Dissipation theorem(FDT). For the same side correlators and different side correlators it looks
slightly different. For the same side case we have:
\beq
\frac{\Im \l G^>_{T,aa}(\omega)+G^<_{T,aa}(\omega) \r}{(-2) \Im G^R_{T,aa}(\omega)} = -\tanh \frac{\beta(T) \omega}{2}
\label{tanh}
\eeq
where for different sides it reads as:
\beq
\frac{\Re \l G^>_{T,ab}(\omega)+G^<_{T,ab}(\omega) \r}{(-2) \Re G^R_{T,ab}(\omega)} = -\tanh \frac{\beta(T) \omega}{2}, 
\ a \neq b
\eeq

To the extent that  the system is in   thermal equilibrium, it is time translation invariant and it does not matter what the value of $t_1 + t_2$ is. We will actually choose the 
  corner slice Green function $G^{>}_{T,ab}$ in \nref{Corner}.
This choice is motivated by the fact that exactly these Greens function enter in the definition of energy (\ref{kb_energy}).

\begin{figure}[!ht]
\centering
\includegraphics[scale=0.7]{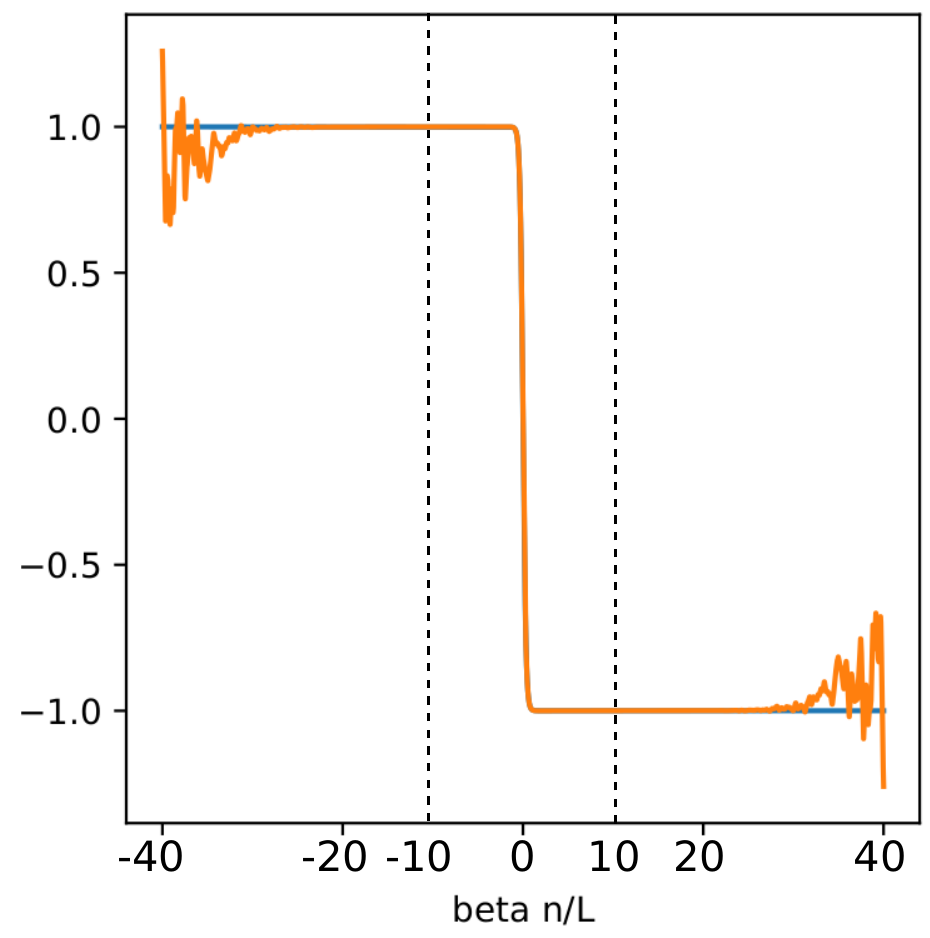}
\minipage{0.8\textwidth}
\centering
\caption{\small (Orange) The left hand side of eq. (\ref{tanh}) for $\beta=80$, $dt=0.2$. The Green function
is defined on a strip of width $L=1600$. (Blue) $tanh$ function. We expect that they should match for frequencies much less than the inverse discretization timestep: $n/L \ll 1/dt$.
In on this graph this UV cutoff is at   $\beta n_{\rm max}/L \sim 400$. The region used for the fit,  $\beta |n|/L \leq 10$, is within the dashed lines.    } 
\label{fig:tanh}
\endminipage
\end{figure}

Then one can fit it the left hand side of \nref{tanh} with the tanh function to extract the temperature. We need to select a frequency range of at least a couple of temperatures 
in order to really probe the thermalization. On the other hand, for very large frequencies the numerical data has a lot of noise coming from discrete timestep, see Figure \ref{fig:tanh}.
Typically we will fit eq. (\ref{tanh}) up to frequencies of order 10 temperatures, i.e. $\beta n/L \sim 10$. Although
the results do not depend much of this choice.

\section{Change of boundary conditions}
\label{app:flow}
First we need derive an analogue of (\ref{NDdet}) for the case of global $AdS_2$. 
We start from fermions having Dirichlet boundary conditions on both boundaries and add an extra fermion $\chi_{L,R}$ on the boundary to initiate the flow from Dirichlet to Neumann boundary
conditions. Essentially this way we perform a Legendre transformation \cite{KlebanovWitten}. This way we obtain the relation:
\beq
Z_N = Z_D \sqrt{\det \hat G_{1-\Delta}}
\eeq
where $\hat{G}_{1-\Delta}$ is the matrix
\beq
\begin{pmatrix}
G^{LL}_{1-\Delta} & G^{LR}_{1-\Delta} \\
G^{RL}_{1-\Delta} & G^{RR}_{1-\Delta} 
\end{pmatrix}
\eeq
with with standard conformal correlators 
\begin{align}
G^{LL}_{1-\Delta} \propto \l \sinh \l \frac{t}{2} \r \r^{2\Delta-2} \\
G^{LR}_{1-\Delta} \propto \l \cosh \l \frac{t}{2} \r \r^{2\Delta-2}
\end{align}

Now let us return to our problem with the $\mu$ term. So now we have Neumann fermions plus the interaction term $\mu \psi_L \psi_R$ coupling the two boundaries.
The partition function can be easily found:
\beq
Z_{\mu} = Z_{\mu=0} \sqrt{\det \l 1 + \hat \mu \hat G_{\Delta} \r}
\eeq
where the matrix $\hat{\mu}$ is given by eq. (\ref{mu_ab}). For large $\mu$ we have
\beq
Z_{\mu \gg 1} \propto Z_N \sqrt{\det \hat{G}_{\Delta}}
\eeq
To conclude that $Z_{\mu \gg 1} \propto Z_D$ we need the matrix relation
\beq
\hat{G}_{1-\Delta}(\omega) \hat{G}_{\Delta}(\omega) = \text{id}
\eeq
In fact this relation coincides with the conformal(i.e. neglecting the time derivative) Dyson--Schwinger equation (\ref{eq:ds}).

\bibliographystyle{apsrev4-1long}
\bibliography{Wormhole.bib}

\begin{thebibliography}{10}%
\makeatletter
\providecommand \@ifxundefined [1]{%
 \ifx #1\undefined \expandafter \@firstoftwo
 \else \expandafter \@secondoftwo
\fi
}%
\providecommand \@ifnum [1]{%
 \ifnum #1\expandafter \@firstoftwo
 \else \expandafter \@secondoftwo
\fi
}%
\providecommand \enquote [1]{``#1''}%
\providecommand \bibnamefont  [1]{#1}%
\providecommand \bibfnamefont [1]{#1}%
\providecommand \citenamefont [1]{#1}%
\providecommand\href[0]{\@sanitize\@href}%
\providecommand\@href[1]{\endgroup\@@startlink{#1}\endgroup\@@href}%
\providecommand\@@href[1]{#1\@@endlink}%
\providecommand \@sanitize [0]{\begingroup\catcode`\&12\catcode`\#12\relax}%
\@ifxundefined \pdfoutput {\@firstoftwo}{%
 \@ifnum{\z@=\pdfoutput}{\@firstoftwo}{\@secondoftwo}%
}{%
 \providecommand\@@startlink[1]{\leavevmode\special{html:<a href="#1">}}%
 \providecommand\@@endlink[0]{\special{html:</a>}}%
}{%
 \providecommand\@@startlink[1]{%
  \leavevmode
  \pdfstartlink
   attr{/Border[0 0 1 ]/H/I/C[0 1 1]}%
   user{/Subtype/Link/A<</Type/Action/S/URI/URI(#1)>>}%
  \relax
 }%
 \providecommand\@@endlink[0]{\pdfendlink}%
}%
\providecommand \url  [0]{\begingroup\@sanitize \@url }%
\providecommand \@url [1]{\endgroup\@href {#1}{\urlprefix}}%
\providecommand \urlprefix [0]{URL }%
\providecommand \Eprint[0]{\href }%
\@ifxundefined \urlstyle {%
  \providecommand \doi [1]{doi:\discretionary{}{}{}#1}%
}{%
  \providecommand \doi [0]{doi:\discretionary{}{}{}\begingroup
  \urlstyle{rm}\Url }%
}%
\providecommand \doibase [0]{http://dx.doi.org/}%
\providecommand \Doi[1]{\href{\doibase#1}}%
\providecommand \bibAnnote [3]{%
  \BibitemShut{#1}%
  \begin{quotation}\noindent
    \textsc{Key:}\ #2\\\textsc{Annotation:}\ #3%
  \end{quotation}%
}%
\providecommand \bibAnnoteFile [2]{%
  \IfFileExists{#2}{\bibAnnote {#1} {#2} {\input{#2}}}{}%
}%
\providecommand \typeout [0]{\immediate \write \m@ne }%
\providecommand \selectlanguage [0]{\@gobble}%
\providecommand \bibinfo [0]{\@secondoftwo}%
\providecommand \bibfield [0]{\@secondoftwo}%
\providecommand \translation [1]{[#1]}%
\providecommand \BibitemOpen[0]{}%
\providecommand \bibitemStop [0]{}%
\providecommand \bibitemNoStop [0]{.\EOS\space}%
\providecommand \EOS [0]{\spacefactor3000\relax}%
\providecommand \BibitemShut [1]{\csname bibitem#1\endcsname}%
\bibitem{SachdevYe}%
  \BibitemOpen
  \bibfield{author}{%
  \bibinfo {author} {\bibfnamefont{Subir}\ \bibnamefont{Sachdev}}\ and\
  \bibinfo {author} {\bibfnamefont{Jinwu}\ \bibnamefont{Ye}},\ }%
  \bibfield{title}{%
  \enquote{\bibinfo {title} {{Gapless spin fluid ground state in a random,
  quantum Heisenberg magnet}},}\ }%
  \bibfield{journal}{%
  \Doi{10.1103/PhysRevLett.70.3339}{\bibinfo {journal} {Phys. Rev. Lett.}}\ }%
  \textbf{\bibinfo {volume} {70}},\ \bibinfo {pages} {3339} (\bibinfo {year}
  {1993}),\
  \Eprint{http://arxiv.org/abs/cond-mat/9212030}{arXiv:cond-mat/9212030
  [cond-mat]}%
  \bibAnnoteFile{NoStop}{SachdevYe}%
\bibitem{kitaev}%
  \BibitemOpen
  \bibfield{author}{%
  \bibinfo {author} {\bibfnamefont{Alexei}\ \bibnamefont{Kitaev}},\ }%
  \bibfield{title}{%
  \enquote{\bibinfo {title} {A simple model of quantum holography.}.}\ }%
  \bibinfo {journal} {Talks at KITP, April 7, 2015 and May 27, 2015.}%
  \bibAnnoteFile{Stop}{kitaev}%
\bibitem{MS}%
  \BibitemOpen
\bibfield{journal}{%
    }%
  \bibfield{author}{%
  \bibinfo {author} {\bibfnamefont{Juan}\ \bibnamefont{Maldacena}}\ and\
  \bibinfo {author} {\bibfnamefont{Douglas}\ \bibnamefont{Stanford}},\ }%
  \bibfield{title}{%
  \enquote{\bibinfo {title} {{Remarks on the Sachdev-Ye-Kitaev model}},}\ }%
  \bibfield{journal}{%
  \Doi{10.1103/PhysRevD.94.106002}{\bibinfo {journal} {Phys. Rev.}}\ }%
  \textbf{\bibinfo {volume} {D94}},\ \bibinfo {pages} {106002} (\bibinfo {year}
  {2016}),\ \Eprint{http://arxiv.org/abs/1604.07818}{arXiv:1604.07818
  [hep-th]}%
  \bibAnnoteFile{NoStop}{MS}%
\bibitem{KitaevSuhFirstPaper}%
  \BibitemOpen
  \bibfield{author}{%
  \bibinfo {author} {\bibfnamefont{Alexei}\ \bibnamefont{Kitaev}}\ and\
  \bibinfo {author} {\bibfnamefont{S.~Josephine}\ \bibnamefont{Suh}},\ }%
  \bibfield{title}{%
  \enquote{\bibinfo {title} {{The soft mode in the Sachdev-Ye-Kitaev model and
  its gravity dual}},}\ }%
  \bibfield{journal}{%
  \Doi{10.1007/JHEP05(2018)183}{\bibinfo {journal} {JHEP}}\ }%
  \textbf{\bibinfo {volume} {05}},\ \bibinfo {pages} {183} (\bibinfo {year}
  {2018}),\ \Eprint{http://arxiv.org/abs/1711.08467}{arXiv:1711.08467
  [hep-th]}%
  \bibAnnoteFile{NoStop}{KitaevSuhFirstPaper}%
\bibitem{Jackiw}%
  \BibitemOpen
  \bibfield{author}{%
  \bibinfo {author} {\bibfnamefont{R.}~\bibnamefont{Jackiw}},\ }%
  \bibfield{title}{%
  \enquote{\bibinfo {title} {{Lower Dimensional Gravity}},}\ }%
  \bibfield{booktitle}{%
  \emph{\bibinfo {booktitle} {{1984 Meeting of the Division of Particles and
  Fields of the APS Santa Fe, New Mexico, October 31-November 3, 1984}}},\ }%
  \bibfield{journal}{%
  \Doi{10.1016/0550-3213(85)90448-1}{\bibinfo {journal} {Nucl. Phys.}}\ }%
  \textbf{\bibinfo {volume} {B252}},\ \bibinfo {pages} {343--356} (\bibinfo
  {year} {1985})%
  \bibAnnoteFile{NoStop}{Jackiw}%
\bibitem{Teitelboim}%
  \BibitemOpen
  \bibfield{author}{%
  \bibinfo {author} {\bibfnamefont{C.}~\bibnamefont{Teitelboim}},\ }%
  \bibfield{title}{%
  \enquote{\bibinfo {title} {{Gravitation and Hamiltonian Structure in Two
  Space-Time Dimensions}},}\ }%
  \bibfield{journal}{%
  \Doi{10.1016/0370-2693(83)90012-6}{\bibinfo {journal} {Phys. Lett.}}\ }%
  \textbf{\bibinfo {volume} {126B}},\ \bibinfo {pages} {41--45} (\bibinfo
  {year} {1983})%
  \bibAnnoteFile{NoStop}{Teitelboim}%
\bibitem{AP}%
  \BibitemOpen
  \bibfield{author}{%
  \bibinfo {author} {\bibfnamefont{Ahmed}\ \bibnamefont{Almheiri}}\ and\
  \bibinfo {author} {\bibfnamefont{Joseph}\ \bibnamefont{Polchinski}},\ }%
  \bibfield{title}{%
  \enquote{\bibinfo {title} {{Models of AdS$_{2}$ backreaction and
  holography}},}\ }%
  \bibfield{journal}{%
  \Doi{10.1007/JHEP11(2015)014}{\bibinfo {journal} {JHEP}}\ }%
  \textbf{\bibinfo {volume} {11}},\ \bibinfo {pages} {014} (\bibinfo {year}
  {2015}),\ \Eprint{http://arxiv.org/abs/1402.6334}{arXiv:1402.6334 [hep-th]}%
  \bibAnnoteFile{NoStop}{AP}%
\bibitem{MQ}%
  \BibitemOpen
  \bibfield{author}{%
  \bibinfo {author} {\bibfnamefont{Juan}\ \bibnamefont{Maldacena}}\ and\
  \bibinfo {author} {\bibfnamefont{Xiao-Liang}\ \bibnamefont{Qi}},\ }%
  \bibfield{title}{%
  \enquote{\bibinfo {title} {{Eternal traversable wormhole}},}\ }%
   (\bibinfo {year} {2018}),\
  \Eprint{http://arxiv.org/abs/1804.00491}{arXiv:1804.00491 [hep-th]}%
  \bibAnnoteFile{NoStop}{MQ}%
\bibitem{Garcia-Garcia:2019poj}%
  \BibitemOpen
  \bibfield{author}{%
  \bibinfo {author} {\bibfnamefont{Antonio~M.}\
  \bibnamefont{Garc\'{\i}a-Garc\'{\i}a}}, \bibinfo {author}
  {\bibfnamefont{Tomoki}\ \bibnamefont{Nosaka}}, \bibinfo {author}
  {\bibfnamefont{Dario}\ \bibnamefont{Rosa}},\ and\ \bibinfo {author}
  {\bibfnamefont{Jacobus J.~M.}\ \bibnamefont{Verbaarschot}},\ }%
  \bibfield{title}{%
  \enquote{\bibinfo {title} {{Quantum chaos transition in a two-site
  Sachdev-Ye-Kitaev model dual to an eternal traversable wormhole}},}\ }%
  \bibfield{journal}{%
  \Doi{10.1103/PhysRevD.100.026002}{\bibinfo {journal} {Phys. Rev.}}\ }%
  \textbf{\bibinfo {volume} {D100}},\ \bibinfo {pages} {026002} (\bibinfo
  {year} {2019}),\ \Eprint{http://arxiv.org/abs/1901.06031}{arXiv:1901.06031
  [hep-th]}%
  \bibAnnoteFile{NoStop}{Garcia-Garcia:2019poj}%
\bibitem{Kim:2019upg}%
  \BibitemOpen
  \bibfield{author}{%
  \bibinfo {author} {\bibfnamefont{Jaewon}\ \bibnamefont{Kim}}, \bibinfo
  {author} {\bibfnamefont{Igor~R.}\ \bibnamefont{Klebanov}}, \bibinfo {author}
  {\bibfnamefont{Grigory}\ \bibnamefont{Tarnopolsky}},\ and\ \bibinfo {author}
  {\bibfnamefont{Wenli}\ \bibnamefont{Zhao}},\ }%
  \bibfield{title}{%
  \enquote{\bibinfo {title} {{Symmetry Breaking in Coupled SYK or Tensor
  Models}},}\ }%
  \bibfield{journal}{%
  \Doi{10.1103/PhysRevX.9.021043}{\bibinfo {journal} {Phys. Rev.}}\ }%
  \textbf{\bibinfo {volume} {X9}},\ \bibinfo {pages} {021043} (\bibinfo {year}
  {2019}),\ \Eprint{http://arxiv.org/abs/1902.02287}{arXiv:1902.02287
  [hep-th]}%
  \bibAnnoteFile{NoStop}{Kim:2019upg}%
\bibitem{GaoJafferisRecent}%
  \BibitemOpen
  \bibfield{author}{%
  \bibinfo {author} {\bibfnamefont{Ping}\ \bibnamefont{Gao}}\ and\ \bibinfo
  {author} {\bibfnamefont{Daniel~Louis}\ \bibnamefont{Jafferis}},\ }%
  \bibfield{title}{%
  \enquote{\bibinfo {title} {{A Traversable Wormhole Teleportation Protocol in
  the SYK Model}},}\ }%
   (\bibinfo {year} {2019}),\
  \Eprint{http://arxiv.org/abs/1911.07416}{arXiv:1911.07416 [hep-th]}%
  \bibAnnoteFile{NoStop}{GaoJafferisRecent}%
\bibitem{MMP}%
  \BibitemOpen
  \bibfield{author}{%
  \bibinfo {author} {\bibfnamefont{Juan}\ \bibnamefont{Maldacena}}, \bibinfo
  {author} {\bibfnamefont{Alexey}\ \bibnamefont{Milekhin}},\ and\ \bibinfo
  {author} {\bibfnamefont{Fedor}\ \bibnamefont{Popov}},\ }%
  \bibfield{title}{%
  \enquote{\bibinfo {title} {{Traversable wormholes in four dimensions}},}\ }%
   (\bibinfo {year} {2018}),\
  \Eprint{http://arxiv.org/abs/1807.04726}{arXiv:1807.04726 [hep-th]}%
  \bibAnnoteFile{NoStop}{MMP}%
\bibitem{GaoJafferisWall}%
  \BibitemOpen
  \bibfield{author}{%
  \bibinfo {author} {\bibfnamefont{Ping}\ \bibnamefont{Gao}}, \bibinfo {author}
  {\bibfnamefont{Daniel~Louis}\ \bibnamefont{Jafferis}},\ and\ \bibinfo
  {author} {\bibfnamefont{Aron~C.}\ \bibnamefont{Wall}},\ }%
  \bibfield{title}{%
  \enquote{\bibinfo {title} {{Traversable Wormholes via a Double Trace
  Deformation}},}\ }%
  \bibfield{journal}{%
  \Doi{10.1007/JHEP12(2017)151}{\bibinfo {journal} {JHEP}}\ }%
  \textbf{\bibinfo {volume} {12}},\ \bibinfo {pages} {151} (\bibinfo {year}
  {2017}),\ \Eprint{http://arxiv.org/abs/1608.05687}{arXiv:1608.05687
  [hep-th]}%
  \bibAnnoteFile{NoStop}{GaoJafferisWall}%
\bibitem{syk_bath}%
  \BibitemOpen
  \bibfield{author}{%
  \bibinfo {author} {\bibfnamefont{Ahmed}\ \bibnamefont{Almheiri}}, \bibinfo
  {author} {\bibfnamefont{Alexey}\ \bibnamefont{Milekhin}},\ and\ \bibinfo
  {author} {\bibfnamefont{Brian}\ \bibnamefont{Swingle}},\ }%
  \bibfield{title}{%
  \enquote{\bibinfo {title} {{Universal Constraints on Energy Flow and SYK
  Thermalization}},}\ }%
   (\bibinfo {month} {12}\ \bibinfo {year} {2019}),\
  \Eprint{http://arxiv.org/abs/1912.04912}{arXiv:1912.04912 [hep-th]}%
  \bibAnnoteFile{NoStop}{syk_bath}%
\bibitem{quench1}%
  \BibitemOpen
  \bibfield{author}{%
  \bibinfo {author} {\bibfnamefont{Andreas}\ \bibnamefont{Eberlein}}, \bibinfo
  {author} {\bibfnamefont{Valentin}\ \bibnamefont{Kasper}}, \bibinfo {author}
  {\bibfnamefont{Subir}\ \bibnamefont{Sachdev}},\ and\ \bibinfo {author}
  {\bibfnamefont{Julia}\ \bibnamefont{Steinberg}},\ }%
  \bibfield{title}{%
  \enquote{\bibinfo {title} {{Quantum quench of the Sachdev-Ye-Kitaev
  Model}},}\ }%
  \bibfield{journal}{%
  \Doi{10.1103/PhysRevB.96.205123}{\bibinfo {journal} {Phys. Rev.}}\ }%
  \textbf{\bibinfo {volume} {B96}},\ \bibinfo {pages} {205123} (\bibinfo {year}
  {2017}),\ \Eprint{http://arxiv.org/abs/1706.07803}{arXiv:1706.07803
  [cond-mat.str-el]}%
  \bibAnnoteFile{NoStop}{quench1}%
\bibitem{quench2}%
  \BibitemOpen
  \bibfield{author}{%
  \bibinfo {author} {\bibfnamefont{Ritabrata}\ \bibnamefont{Bhattacharya}},
  \bibinfo {author} {\bibfnamefont{Dileep~P.}\ \bibnamefont{Jatkar}},\ and\
  \bibinfo {author} {\bibfnamefont{Nilakash}\ \bibnamefont{Sorokhaibam}},\ }%
  \bibfield{title}{%
  \enquote{\bibinfo {title} {{Quantum Quenches and Thermalization in SYK
  models}},}\ }%
   (\bibinfo {year} {2018}),\
  \Eprint{http://arxiv.org/abs/1811.06006}{arXiv:1811.06006 [hep-th]}%
  \bibAnnoteFile{NoStop}{quench2}%
\bibitem{cft_breaking}%
  \BibitemOpen
  \bibfield{author}{%
  \bibinfo {author} {\bibfnamefont{Juan}\ \bibnamefont{Maldacena}}, \bibinfo
  {author} {\bibfnamefont{Douglas}\ \bibnamefont{Stanford}},\ and\ \bibinfo
  {author} {\bibfnamefont{Zhenbin}\ \bibnamefont{Yang}},\ }%
  \bibfield{title}{%
  \enquote{\bibinfo {title} {{Conformal symmetry and its breaking in two
  dimensional Nearly Anti-de-Sitter space}},}\ }%
  \bibfield{journal}{%
  \Doi{10.1093/ptep/ptw124}{\bibinfo {journal} {PTEP}}\ }%
  \textbf{\bibinfo {volume} {2016}},\ \bibinfo {pages} {12C104} (\bibinfo
  {year} {2016}),\ \Eprint{http://arxiv.org/abs/1606.01857}{arXiv:1606.01857
  [hep-th]}%
  \bibAnnoteFile{NoStop}{cft_breaking}%
\bibitem{Jensen:2016pah}%
  \BibitemOpen
  \bibfield{author}{%
  \bibinfo {author} {\bibfnamefont{Kristan}\ \bibnamefont{Jensen}},\ }%
  \bibfield{title}{%
  \enquote{\bibinfo {title} {{Chaos in AdS$_2$ Holography}},}\ }%
  \bibfield{journal}{%
  \Doi{10.1103/PhysRevLett.117.111601}{\bibinfo {journal} {Phys. Rev. Lett.}}\
  }%
  \textbf{\bibinfo {volume} {117}},\ \bibinfo {pages} {111601} (\bibinfo {year}
  {2016}),\ \Eprint{http://arxiv.org/abs/1605.06098}{arXiv:1605.06098
  [hep-th]}%
  \bibAnnoteFile{NoStop}{Jensen:2016pah}%
\bibitem{Engelsoy:2016xyb}%
  \BibitemOpen
  \bibfield{author}{%
  \bibinfo {author} {\bibfnamefont{Julius}\ \bibnamefont{Engelsöy}}, \bibinfo
  {author} {\bibfnamefont{Thomas~G.}\ \bibnamefont{Mertens}},\ and\ \bibinfo
  {author} {\bibfnamefont{Herman}\ \bibnamefont{Verlinde}},\ }%
  \bibfield{title}{%
  \enquote{\bibinfo {title} {{An investigation of AdS$_{2}$ backreaction and
  holography}},}\ }%
  \bibfield{journal}{%
  \Doi{10.1007/JHEP07(2016)139}{\bibinfo {journal} {JHEP}}\ }%
  \textbf{\bibinfo {volume} {07}},\ \bibinfo {pages} {139} (\bibinfo {year}
  {2016}),\ \Eprint{http://arxiv.org/abs/1606.03438}{arXiv:1606.03438
  [hep-th]}%
  \bibAnnoteFile{NoStop}{Engelsoy:2016xyb}%
\bibitem{KlebanovWitten}%
  \BibitemOpen
  \bibfield{author}{%
  \bibinfo {author} {\bibfnamefont{Igor~R.}\ \bibnamefont{Klebanov}}\ and\
  \bibinfo {author} {\bibfnamefont{Edward}\ \bibnamefont{Witten}},\ }%
  \bibfield{title}{%
  \enquote{\bibinfo {title} {{AdS / CFT correspondence and symmetry
  breaking}},}\ }%
  \bibfield{journal}{%
  \Doi{10.1016/S0550-3213(99)00387-9}{\bibinfo {journal} {Nucl. Phys.}}\ }%
  \textbf{\bibinfo {volume} {B556}},\ \bibinfo {pages} {89--114} (\bibinfo
  {year} {1999}),\
  \Eprint{http://arxiv.org/abs/hep-th/9905104}{arXiv:hep-th/9905104 [hep-th]}%
  \bibAnnoteFile{NoStop}{KlebanovWitten}%
\bibitem{KitaevRecent}%
  \BibitemOpen
  \bibfield{author}{%
  \bibinfo {author} {\bibfnamefont{Yingfei}\ \bibnamefont{Gu}}, \bibinfo
  {author} {\bibfnamefont{Alexei}\ \bibnamefont{Kitaev}}, \bibinfo {author}
  {\bibfnamefont{Subir}\ \bibnamefont{Sachdev}},\ and\ \bibinfo {author}
  {\bibfnamefont{Grigory}\ \bibnamefont{Tarnopolsky}},\ }%
  \bibfield{title}{%
  \enquote{\bibinfo {title} {{Notes on the complex Sachdev-Ye-Kitaev model}},}\
  }%
   (\bibinfo {year} {2019}),\
  \Eprint{http://arxiv.org/abs/1910.14099}{arXiv:1910.14099 [hep-th]}%
  \bibAnnoteFile{NoStop}{KitaevRecent}%
\bibitem{kamenev_book}%
  \BibitemOpen
  \bibfield{author}{%
  \bibinfo {author} {\bibfnamefont{Alex}\ \bibnamefont{Kamenev}},\ }%
  \emph{\bibinfo {title} {Field theory of non-equilibrium systems}}\ (\bibinfo
  {publisher} {Cambridge University Press},\ \bibinfo {year} {2011})%
  \bibAnnoteFile{NoStop}{kamenev_book}%
\bibitem{stefanucci2013}%
  \BibitemOpen
  \bibfield{author}{%
  \bibinfo {author} {\bibfnamefont{Gianluca}\ \bibnamefont{Stefanucci}}\ and\
  \bibinfo {author} {\bibfnamefont{Robert}\ \bibnamefont{Van~Leeuwen}},\ }%
  \emph{\bibinfo {title} {Nonequilibrium many-body theory of quantum systems: a
  modern introduction}}\ (\bibinfo {publisher} {Cambridge University Press},\
  \bibinfo {year} {2013})%
  \bibAnnoteFile{NoStop}{stefanucci2013}%
\bibitem{Maldacena:2019cbz}%
  \BibitemOpen
  \bibfield{author}{%
  \bibinfo {author} {\bibfnamefont{Juan}\ \bibnamefont{Maldacena}}, \bibinfo
  {author} {\bibfnamefont{Gustavo~J.}\ \bibnamefont{Turiaci}},\ and\ \bibinfo
  {author} {\bibfnamefont{Zhenbin}\ \bibnamefont{Yang}},\ }%
  \bibfield{title}{%
  \enquote{\bibinfo {title} {{Two dimensional Nearly de Sitter gravity}},}\ }%
   (\bibinfo {year} {2019}),\
  \Eprint{http://arxiv.org/abs/1904.01911}{arXiv:1904.01911 [hep-th]}%
  \bibAnnoteFile{NoStop}{Maldacena:2019cbz}%
\bibitem{Chen_2018}%
  \BibitemOpen
  \bibfield{author}{%
  \bibinfo {author} {\bibfnamefont{Anffany}\ \bibnamefont{Chen}}, \bibinfo
  {author} {\bibfnamefont{R}~\bibnamefont{Ilan}}, \bibinfo {author}
  {\bibfnamefont{F}~\bibnamefont{de~Juan}}, \bibinfo {author}
  {\bibfnamefont{DI}~\bibnamefont{Pikulin}},\ and\ \bibinfo {author}
  {\bibfnamefont{M}~\bibnamefont{Franz}},\ }%
  \bibfield{title}{%
  \enquote{\bibinfo {title} {Quantum holography in a graphene flake with an
  irregular boundary},}\ }%
  \bibfield{journal}{%
  \bibinfo {journal} {Physical review letters}\ }%
  \textbf{\bibinfo {volume} {121}},\ \bibinfo {pages} {036403} (\bibinfo {year}
  {2018})%
  \bibAnnoteFile{NoStop}{Chen_2018}%
\end{thebibliography}%
\end{document}